# SPECIFICATION, CONSTRUCTION, AND EXACT REDUCTION OF STATE TRANSITION SYSTEM MODELS OF BIOCHEMICAL PROCESSES

**Scott M. Bugenhagen and Daniel A. Beard**

**Department of Physiology, Medical College of Wisconsin, Milwaukee, WI 53226**

**Corresponding author:** DA Beard, dbeard@mcw.edu

**Abstract**

Biochemical reaction systems may be viewed as discrete event processes characterized by a number of states and state transitions. These systems may be modeled as state transition systems with transitions representing individual reaction events. Since they often involve a large number of interactions, it can be difficult to construct such a model for a system, and since the resulting state-level model can involve a huge number of states, model analysis can be difficult or impossible. Here, we describe methods for the high-level specification of a system using hypergraphs, for the automated generation of a state-level model from a high-level model, and for the exact reduction of a state-level model using information from the high-level model. Exact reduction is achieved through the automated application of symmetry reduction and invariant manifold reduction techniques to the high-level model, allowing potentially significant reductions without the need to generate a full model. The application of the method to biochemical reaction systems is illustrated by models describing a hypothetical ion-channel at several levels of complexity. The method allows for the reduction of the otherwise intractable example models to a manageable size.

## I. Introduction

A system model aims to predict the integrated behavior of a number of interacting system components. One class of system with particular relevance to biochemical processes is the discrete-state continuous-time system, where bimolecular association/dissociation reactions and unimolecular conformational changes are represented as discrete events representing the elementary reactions of biochemical systems. With this view, a biochemical network may be represented as a set of discrete reaction events occurring between a set of molecules. The qualitative behavior of a biochemical network may then be modeled as a labeled transition system (LTS), consisting of a set of states and labeled transitions between the states, where the transitions correspond to biochemical reaction events. Additionally, the quantitative behavior may be modeled as a continuous-time Markov chain, which is simply a special case of an LTS where transitions are labeled with rate constants.

A major drawback of modeling a biochemical network as an LTS (or as a CTMC) is the well known state-space explosion problem - a rapid (exponential) growth in the state space of a system with the numbers of molecules and reactions in the system. As the state space of a system becomes large, the process of manually constructing the state-level model (i.e. the LTS or CTMC) for the system becomes tedious and error prone, and memory requirements for storing the state-level model become excessive. For this reason, techniques for rule-based modeling have garnered increasing interest amongst the biochemical modeling community as evidenced by the release of several tools designed for this purpose[1,2,3,4]. The basic approach used by many of these tools to deal with exploding state-spaces is to define a high-level model (HLM) describing the reaction rules according to some high-level specification method (HLSM), and to use the HLM along with a kinetic Monte Carlo (KMC) algorithm to simulate sample trajectories of a reaction. Using this approach, states are generated "on-the-fly," thereby avoiding the need to generate a state-level model.

While the simulation approach (KMC on an HLM) works well for a wide-variety of biochemical systems, there remain certain classes of systems for which this approach leads to significant difficulty. Some examples include: 1) stiff systems, defined as those containing transition rates differing by orders of magnitude; 2) biochemical systems associated with rare events; and 3) multiscale systems integrating biochemical reactions and reaction networks with physiological processes at the cell, tissue, organ, and whole body levels. There exist variations on standard KMC algorithms that are better able to cope with stiff systems[5], however, the statistical nature of KMC and its reliance on confidence intervals is a fundamental limitation with regard to its ability to compute bounds on measures associated with rare events. (Since the width of a confidence interval scales by the inverse of the square-root of the sample size, achieving narrow enough confidence intervals on rare events can require huge sample sizes.) In addition, simulating a biochemical network in the context of a multiscale physiological system almost always requires integration with a global system of ODEs and/or PDEs, and it is usually the case that many instances of a single biochemical network occur simultaneously at the physiological scale of interest. For example, a cell may contain thousands of receptor complexes on its membrane, and a tissue or organ may be composed of millions of cells, each expressing a variable number of copies of a gene product associated with a gene regulatory network. Thus, stochastic simulation of a physiological process at a particular scale of interest potentially requires a large number of reaction trajectories per sample run, and therefore is generally not a suitable approach for use in multiscale modeling applications.

Although a large focus on the rule-based modeling of biochemical systems has been geared toward simulation using KMC algorithms, there exists a variety of powerful numerical approaches for the analysis of a HLM that have been developed and applied to many different problems (particularly in the field of operations) over the last few decades. These approaches may be divided into two major classes: 1) those designed to *tolerate* largeness; and 2) those designed to *avoid* largeness. Approaches to tolerate largeness are mainly centered upon use of symbolic data structures and associated algorithms[6], whereas approaches to avoid largeness are based on either using a HLM to generate only a part of the state-level model (e.g. truncation[7] and/or probabilistic evaluation[8,9]), or on using a HLM to identify structure in the associated state-level model (e.g. lumpability[10], invariant manifolds[11], product form solutions[12], etc.). A major advantage of all these approaches over stochastic simulation is that they allow direct computation of measures of interest (subject to an approximation error) such that rare events are much less of an issue. In addition, since the probability distribution of the states of a system is computed directly, incorporating a biochemical network into a multiscale physiological model is straightforward.

Here, we present an approach for largeness avoidance based on model structure identification from a HLM. Specifically, we present a formal description of a HLSM for models of biochemical systems composed of multiple interacting components, we describe how a HLM specified according to our method may be used to identify symmetries and invariant manifolds in a system, and we describe methods for using the identified structure to construct a reduced LTS describing the system. Finally, we illustrate the application of the method with examples.

## II. System Specification and Generation

### *State-Level Model*

As described in the introduction, a biochemical reaction may be viewed as a discrete event occurring in a molecule or between sets of molecules such that the possible behaviors of a system may be described completely by a set of discrete states and transitions between states. Thus, the qualitative behavior of a biochemical reaction system may be described by a directed graph called an LTS, defined as:

$$\mathcal{L} = (F, R, \Delta)$$

where:

1. $F$ is a set of system states,
2. $R$ is a set of transition labels, and
3. $\Delta \subseteq F \times R \times F$ is the transition relation.

An example of a simple LTS is shown in figure 1.

In addition, quantitative measures on a biochemical system may be obtained by mapping $\mathcal{L}$ to a homogeneous CTMC using a weighting function $W_h : R \to \mathbb{R}_{\geq 0}$ in the case of constant transition rates, or to an inhomogeneous CTMC using a weighting function $W_i : R \to \mathbb{R}^t_{\geq 0}$ in the more general case of time-varying rate constants ($\mathbb{R}^t_{\geq 0}$ denotes the set of positive real-valued functions of time $t$). The resulting CTMC may be used to derive the master equation describing the time-evolution of the probabilities of the system occupying each of the states in $F$. This is done by imposing some order on the system states using an indexing function $J$ such that $J : I \to F$, where $I = \{1, 2, ..., |F|\}$ is the index set. The master equation may then be written as

$$\frac{dP}{dt} = AP, \tag{1}$$

where $P = [P_i]$ is a column vector defined such that $P_i$, under the restriction that $P_i > 0$ and $\sum_i P_i = 1$, is the probability of the system occupying state $J(i)$ at time $t$, and $A = [A_{ij}]$ is a matrix of transition rates given by:

$$A_{ij} = \begin{cases} W(r,t) & \text{for } J(j) \xrightarrow{r} J(i) \in \Delta, \\ -\sum_{i \neq j} A_{ij} & \text{for } i = j, \text{ and} \\ 0 & \text{otherwise,} \end{cases} \tag{2}$$

where $W(r,t)$ is the weight (transition rate) associated with label $r \in R$ at time $t$.

The LTS and associated CTMC are analytical (state-level) representations of a system that may be used directly to analyze its behavior; however they are far too cumbersome to be of much value in system specification. For that, we rely on a high-level representation of the system as described below.

***High-Level Model***

A HLSM allows specification of complex models in a compact and intuitive form with the purpose of simplifying the modeling process and avoiding modeling errors. A HLM should contain all the information required to automatically generate a state-level model of the system. There exist many different classes of HLSMs for the specification of a LTS, each with different advantages, disadvantages, and suitability toward a particular application[13]. Our approach may best be classified as a method based on communicating processes in that it formulates a system as a set of components (processes) that may interact via both synchronizing transitions (transitions occurring simultaneously among a subset of components) and/or functional transitions (transitions in a subset of components with rates that depend on the state of disjoint subsets of components). These types of interactions are common in many biochemical processes, especially in signaling complexes, enzymes, ion channels, etc.

*Component State Space Graph*
The first step of specifying a biochemical reaction system using our approach is to specify the components. This is done using a structure called the component state space graph $\mathcal{C}$, defined as:

$$\mathcal{C} = (S, Q)$$

where:

1. $S = \{s_1, s_2, ..., s_n\}$ is a set of $n$ component states,
2. $Q = \{Q_1, Q_2, ... Q_m\}$, with $Q_j = (X_j, Y_j, Z_j)$ for $j = 1, ..., m$, is a set of component transitions where each $X_j \subseteq S$ is a set of source states, each $Y_j$ is a mapping $Y_j : X_j \to S$ of source states to destination states, and each $Z_j$ is a mapping $Z_j : X_j \to \mathbb{N}$ such that $Z_j(s_i)$ is the multiplicity, or number of occurrences, of components in state $s_i$ required to enable transition $Q_j$.

(Here, the notation
$X_j^{(1)}, X_j^{(2)}, ..., X_j^{(l \le n)} \xrightarrow{Z_j\left(X_j^{(1)}\right), Z_j\left(X_j^{(2)}\right), ..., Z_j\left(X_j^{(l \le n)}\right)} Y_j\left(X_j^{(1)}\right), Y_j\left(X_j^{(2)}\right), ..., Y_j\left(X_j^{(l \le n)}\right)$ is used to represent a component transition $Q_j$.) An important aspect of the above definition is that it allows specification of synchronized transitions. Any component transition $Q_j$ with more than one element in $X_j$ or with $Z_j(s_i) > 1$ for any $s_i \in X_j$ is a synchronizing transition.

Each component state in $S$ may be associated with a component type by finding a partition $\{S_1, S_2, ... S_r\} \vdash S$ of $S$ into $r$ maximal connected subsets and defining a function $\mathcal{T} : S \to \{1, 2, ..., r\}$ such that $\mathcal{T}(s) = i$ for all $s \in S_i$. Then, $\mathcal{T}(s)$ is the component type associated with component state $s \in S$. The usefulness of this definition is described below. (The practical application of these definitions is illustrated below in the section *Simple Example*).

*System Domain*
The next step in system specification is to define the numbers and positions of system components. The system domain $D = \{d_1, d_2, ..., d_p\}$ is an indexed set of $p$ component positions or places. The state of the system is described by a mapping of each element (position) in $D$ to a component state in $S$. That is, the system state is a function $f$ from $D$ to $S$. (For convenience, $f$ may be represented as a $p$-tuple.) The type of component assigned to each position $d_i \in D$ is given by $\mathcal{T}\left(f_{init}(d_i)\right)$, where $f_{init}$ is an initial system state. In general, the set $F$ of system states reachable from some initial system state $f_{init}$ is a proper subset of $S^D$ (where $S^D = \{f \mid f : D \to S\}$) such that the subset of possible component states assigned to $d_i$ is restricted to $S_{\mathcal{T}(f_{init}(d_i))}$. $F$ may be further restricted by the existence of synchronizing and/or functional interactions.

*Dependency Graph*

The final step in system specification is to define the functional dependencies of component transitions; in other words, to define a set of rules for mapping a particular component transition to a transition rate. These rules are given in the form of a directed hypergraph with labeled hyperarcs, which we refer to as the dependency graph $\mathcal{D}(\alpha)$ of a transition rate $\alpha$, defined as:

$$\mathcal{D}(\alpha) = (D, E, L)$$

where:

1. $D = \{d_1, d_2, ..., d_p\}$ is the system domain,
2. $E = \{E_1, E_2, ..., E_q\}$, with $E_y = (A_y, B_y)$ for $y = 1, ..., q$, is a set of hyperarcs with tail sets $A_y \in \mathcal{P}(D)$ and head sets $B_y \in \mathcal{P}(D)$, where $\mathcal{P}(D)$ is the power set of $D$.
3. $L = \{L_y \in L \mid L_y : B_y \to \mathcal{P}(S)\}$ for $y = 1, ..., q$, is a set of labeling functions mapping elements of the head sets of each $E_y$ to elements (subsets of component states) in the power set of $S$.

The hyperarcs $E_y \in E$ represent dependencies of transitions in components pointed to by tail sets $A_y$ on the state of components pointed to by head sets $B_y$. All $A_y$ are singleton for local (non-synchronizing) transitions. Synchronizing transitions have $E$ with each $A_y$ containing $d_x$ for each component involved in the transition. Each element $d_x$ in each $B_y$ is labeled according to a subset of states required to enable a transition in the components of $A_y$ at a specific rate constant $\alpha$. Two $E_y \in E$ may have identical tail sets if there is more than one condition under which a specific rate constant applies for a particular $A_y$. The set $E$ is empty if there are no functional dependencies of the corresponding rate constant.

***Generating the State-Level Model***

The HLM of a system is used to generate the state-space graph $\mathcal{S}$ of a system, which is a modified LTS describing the set of all system states $f \in S^D$ that are reachable from an initial system state $f_{init}$ (i.e., it is the reachability graph spanning the set of all $f \in S^D$ such that there exists a path starting from $f_{init}$ and ending at $f$). $\mathcal{S}$ is defined as follows:

$$\mathcal{S} = (D, F, R, \Delta, \mathcal{C}, \mathcal{D}, f_{init})$$

where:

1. $D = \{d_1, d_2, ...d_p\}$ is the system domain,

2. $F \subseteq S^D$, with $S^D = \{ f \mid f : D \rightarrow S \}$, is a set of system states,
3. $R = \{ R_1, R_2, ..., R_m \}$ is a family of sets of transition labels, each $R_j$ being associated with a component transition $Q_j$,
4. $\Delta \subseteq F \times \bigcup_j R_j \times F$ is the transition relation,
5. $\mathcal{C}$ is the component state-space graph,
6. $\mathcal{D} = \left\{ \mathcal{D}\left(r_k^{(j)}\right) \mid r_k^{(j)} \in \bigcup_j R_j \right\}$ is the set of all dependency graphs, and
7. $f_{init} \in F$ is an initial system state.

An algorithm (modified from Algorithm 2.1 in Ref. 14 to account for the HLM) for generating the state-space graph $\mathcal{S}$ is given below:

**Algorithm 1.** An algorithm for generating the full state-space graph.

1: Set $unprocessed = \{ f_{init} \}$
2: Set $F = \{ f_{init} \}$
3: Set $\Delta = \emptyset$
4: **while** $unprocessed \neq \emptyset$ **do**
5: Select any $f \in unprocessed$ and set $unprocessed = unprocessed \setminus \{ f \}$
6: **for all** $Q_j \in \left\{ Q_j \in Q \mid \left( X_j, Z_j \right) \subseteq f\left[ D \right] \right\}$ **do**
7: **for all** $D' \in \left\{ D' \in \mathcal{P}\left( D \right) \mid f\left[ D' \right] = \left( X_j, Z_j \right) \right\}$ **do**
8: Set $f' = f$
9: **for all** $\mathcal{D}\left( r_k^{(j)} \right) \in \left\{ \mathcal{D}\left( r_k^{(j)} \right) \in \mathcal{D} \mid r_k^{(j)} \in R_j \right\}$ **do**
10: **if** $E = \emptyset$ **then goto** 13
11: **for all** $E_y \in \left\{ E_y \in E \mid A_y = D' \right\}$ **do**
12: **if** $f\left( d \right) \in L_y\left( d \right) \ \forall \ d \in B_y$ **then**
13: **for all** $d \in D'$ **do**
14: Set $f'\left( d \right) = Y_j\left( f\left( d \right) \right)$
15: Set $\Delta = \Delta \cup \left\{ \left( f, r_k^{(j)}, f' \right) \right\}$
16: **if** $f' \notin F$ **then**
17: Set $unprocessed = unprocessed \cup \{ f' \}$
18: Set $F = F \cup \{ f' \}$
19: **return** $\mathcal{S} = \left( D, F, R, \Delta, \mathcal{C}, \mathcal{D}, f_{init} \right)$

In line 6 and 7 of the above algorithm, the notation $f\left[D\right]$ and $f\left[D'\right]$ denote the multiset image (rather than the usual set image) of $D$ and $D'$, respectively, under *f*. (A multiset is a pair $\left(A,m\right)$ where $A$ is a set, or underlying set of elements, and $m: A \to \mathbb{N}$ is a multiplicity function mapping elements in $A$ to the number of occurrences of those elements in the multiset.) The relations in lines 6 and 7 compare these multiset images with the multiset $(X_j, Z_j)$. In the sequel, the image $f\left[A\right]$ of a set $A$ under a function $f$ should be understood as the multiset image when given in the context of a relation involving a multiset, and as the usual set image otherwise.

$\mathcal{S}$ may be mapped to a CTMC by assigning rates to the transition labels using a weighting function as described above for a general LTS.

***Simple Example***

To illustrate the high-level specification of a simple biochemical system using our method, consider a reaction volume containing three molecules, two labeled A and one labeled B, as depicted in Figure 2A. Molecules A can be in one of two conformational states, folded or unfolded, and when in the folded conformation can bind to molecule B. Molecule B has only one conformation and binds to folded molecules A. The possible molecular states and state transitions (reaction events) of all molecules (A and B) in this reaction volume are given by the component state space graph $\mathcal{C}$ illustrated in Figure 2B with transitions tabulated in Table 1. There, the components are defined as molecule A and B binding sites and molecule A conformation. Because, there are two molecules A and one molecule B in the system, the system domain must contain five elements (two molecule A binding sites, two molecule A conformations, and one molecule B binding site), and each element in the domain is assigned to an appropriate component type by the initial state $f_{init}$.

This system contains functional transitions and therefore requires rules for the mapping of those transitions to transition rates in the form of dependency graphs. Specifically, the unfolding of a molecule A ($s_2 \xrightarrow{1} s_1$) at rate $r_1^{(2)}$ requires that it is unbound, and the binding of a molecule A to B ($s_3, s_5 \xrightarrow{1,1} s_4, s_6$) at rate $r_1^{(3)}$ requires that A is in the folded conformation. The dependency graphs $\mathcal{D}\left(r_1^{(2)}\right)$ and $\mathcal{D}\left(r_1^{(3)}\right)$ for the rate constants $r_1^{(2)}$ and $r_1^{(3)}$ are shown in Figure 3 and tabulated in Table 2. (In this system, it is possible to define a different component state space graph with courser components so that the functional transitions described above become non-functional, or constant transitions.) The component state space graph, system domain, dependency graphs, and initial state completely describe the HLM for this simple system, and may be used to generate a state-space graph $\mathcal{S}$ using Algorithm 1. The generated graph is shown in Figure 4A.

## III. Exact Reduction Methods

***Bisimilarity***

A bisimulation of an LTS $\mathcal{L} = (F, R, \Delta)$ is an equivalence relation ~ over $F$ such that for any $f_1, f_2 \in F$, $f_1 \sim f_2$ implies that $f_1$ and $f_2$ behave identically. In this case, $f_1$ and $f_2$ are said to be bisimilar.

There are two principle forms of bisimulation on an LTS, namely forward and backward bisimulation[15]. Forward bisimulation requires equivalence of *outgoing* transitions. More specifically, letting $\Lambda(f_1, f_2)$ be the set of transition labels associated with all transitions from $f_1$ to $f_2$ in Δ, $\sim_f$ is a forward bisimulation if for all $f_1, f_2 \in F$ such that $f_1 \sim_f f_2$, and for all equivalence classes $C \in F/\sim_f$, the following holds:

$$\biguplus_{f' \in C} \Lambda(f_1, f') = \biguplus_{f' \in C} \Lambda(f_2, f'),$$

where $\uplus$ denotes the multiset union. (The multiset union of two multisets $(A,m_A)$ and $(B,m_B)$ is the multiset $(A \cup B, m_A + m_B)$.) This condition implies ordinary, or strong, lumpability of any CTMC to which the LTS is mapped.

By contrast, backward bisimulation requires equivalence of *incoming* transitions. More specifically, $\sim_b$ is a backward bisimulation if for all $f_1, f_2 \in F$ such that $f_1 \sim_b f_2$, and for all equivalence classes $C \in F/\sim_b$, the following holds:

1. $\biguplus_{f' \in C} \Lambda(f', f_1) = \biguplus_{f' \in C} \Lambda(f', f_2)$, and
2. $\biguplus_{f' \in F} \Lambda(f_1, f') = \biguplus_{f' \in F} \Lambda(f_2, f')$.

Condition 2 above requires an equivalent multiset of exit transitions of backward bisimilar states (note that this condition is implicit in the definition of forward bisimulation). Condition 1 and 2 together imply exact lumpability of any CTMC to which the LTS is mapped.

An efficient algorithm exists for computing the coarsest forward (or backward) bisimulation of an LTS[16]; however, this algorithm operates on the full state-space and therefore cannot be used to reduce a system *a priori* (i.e., before state-space generation). It is possible to compute a bisimulation directly from a HLM by considering states that are both forward and backward bisimilar. This is the basis of the symmetry reduction method described below.

***Symmetry Reduction Method***

The symmetry reduction technique is a method for the state-space reduction of a system that works by exploiting symmetries in the system. A complete presentation and detailed analysis of the method, along with a set of algorithms and applications to various system description formalisms, can be found in Junttila's doctoral thesis[14]. Here, the necessary theory behind the method is briefly reviewed and then applied to the system description formalism developed above. The reader is referred to Ref. 14 for a more complete treatment of the subject including proofs.

A symmetry is an automorphism of the state-space graph $\mathcal{S}$. More formally, it is a permutation $\pi$ of the set of system states $F$ such that, for any $f, f' \in F$ and any $r \in \bigcup_j R_j$, the following holds:

$$f \xrightarrow{r} f' \in \Delta \Leftrightarrow \pi(f) \xrightarrow{r} \pi(f') \in \Delta.$$

In other words, a symmetry is a permutation of $F$ which preserves $\Delta$. Since symmetries are permutations, it follows that the set of all symmetries of a state-space graph $\mathcal{S}$ forms an automorphism group, denoted $\mathrm{Aut}(\mathcal{S})$, under the function composition operation. $\mathrm{Aut}(\mathcal{S})$ induces an equivalence relation $\sim$ in $F$ such that for any two states $f_1, f_2 \in F$, $f_1 \sim f_2$ implies that $\pi(f_1) = f_2$ for some $\pi \in \mathrm{Aut}(\mathcal{S})$. Furthermore, $f_1 \sim f_2$ implies both forward and backward bisimilarity of $f_1$ and $f_2$.

It is possible to find $\mathrm{Aut}(\mathcal{S})$ without generating the full state-space graph (which may be impractical or impossible for large models). This is done by identifying a system description level group $G$ of symmetries from the HLM along with a group action $h : G \to \mathrm{Sym}(F)$ of $G$ on $F$ satisfying:

$$h(g) = \pi \Rightarrow f(g(d)) = \pi(f)(d)$$

for all $g \in G$, $\pi \in \mathrm{Aut}(\mathcal{S})$, $f \in F$ and $d \in D$. $G$ can be readily identified from the set of dependency graphs $\mathcal{D}$ according to the following procedure:

1. For each dependency graph $\mathcal{D}\left(r_k^{(j)}\right)$, assign a unique label (e.g. $r_k^{(j)}$) to its hyperarcs.
2. Define a new graph, called the system characteristic graph, as $\mathcal{G} = (V, U)$, where:
   i. $V = D$, and
   ii. $U = \bigcup_{\mathcal{D}\left(r_k^{(j)}\right) \in \mathcal{D}} E_k^{(j)}$.
3. Label each $d \in V$ by $\mathcal{T}(f_{init}(d))$.
4. Find the automorphism group of $\mathcal{G}$.

The characteristic graph $\mathcal{G}$ of the example shown in Figure 2 is illustrated in Figure 5. $\mathrm{Aut}(\mathcal{G})$ can be computed using any general purpose graph automorphism group tool (in our implementation, we use the NAUTY[17] tool). However, most available general-purpose tools require graphs with simple uncolored edges (though vertex colors are usually allowed). A procedure for converting a characteristic graph $\mathcal{G}$ to an equivalent simple (directed) graph $\widetilde{\mathcal{G}}$ is developed in Appendix A1.

Once $G$ is found, it can be used to generate a reduced state-space graph $\widetilde{\mathcal{S}}$, defined as:

$$\widetilde{\mathcal{S}} = \left(D, \widetilde{F}, R, \widetilde{\Delta}, M, \mathcal{C}, \mathcal{D}, f_{init}\right)$$

where:

1. $D = \left\{d_1, d_2, ...d_p\right\}$ as defined in $\mathcal{S}$,
2. $\widetilde{F}$ is a transversal of the quotient set $F/G$,
3. $R = \left\{R_1, R_2, ..., R_m\right\}$ is a family of sets of labels as defined in $\mathcal{S}$,
4. $\widetilde{\Delta} \subseteq \widetilde{F} \times \bigcup_j R_j \times \widetilde{F}$ is the transition relation,
5. $M : \widetilde{\Delta} \to \mathbb{N}$ is a multiplicity function,
6. $\mathcal{C}$ is the component state-space graph,
7. $\mathcal{D} = \left\{\mathcal{D}\left(r_k^{(j)}\right) \mid r_k^{(j)} \in \bigcup_j R_j\right\}$ is the set of all dependency graphs, and
8. $f_{init} \in \widetilde{F}$ is an initial system state.

Note the changes in the definitions of $\widetilde{F}$ and $\widetilde{\Delta}$ and the introduction of the function *M*. The set of system states $\widetilde{F}$ is now a transversal (more specifically, a system of distinct representatives) of the quotient set *F*/*G*, and transitions $\widetilde{\Delta}$ now occur between equivalence class representatives in $\widetilde{F}$. *M* becomes necessary since multiple combinations of components can participate in a single transition in $\widetilde{\Delta}$ (whereas only a single combination of components participated in a transition in Δ).

An algorithm (based on Algorithm 2.2 in Ref. 14) for generating the reduced state-space graph $\widetilde{\mathcal{S}}$ of a system is given below:

**Algorithm 2.** An algorithm for generating the reduced state-space graph

1: Set $unprocessed = \{f_{init}\}$

2: Set $\widetilde{F} = \{f_{init}\}$

3: Set $\widetilde{\Delta} = \varnothing$

4: **while** $unprocessed \neq \varnothing$ **do**

5: Select any $f \in unprocessed$ and set $unprocessed = unprocessed \setminus \{f\}$

6: **for all** $Q_j \in \left\{Q_j \in Q \mid \left(X_j, Z_j\right) \subseteq f[D]\right\}$ **do**

7: **for all** $D' \in \left\{D' \in \mathcal{P}(D) \mid f[D'] = \left(X_j, Z_j\right)\right\}$ **do**

8: Set $f' = f$

9: **for all** $\mathcal{D}\left(r_k^{(j)}\right) \in \left\{\mathcal{D}\left(r_k^{(j)}\right) \in \mathcal{D} \mid r_k^{(j)} \in R_j\right\}$ **do**

10: **if** $E = \varnothing$ **then goto** 13

11: **for all** $E_y \in \left\{E_y \in E \mid A_y = f[D']\right\}$ **do**

12: **if** $f(d) \in L_y(d)\ \forall\ d \in B_y$ **then**

13: **for all** $d \in D'$ **do**

14: Set $f'(d) = Y_j\left(f(d)\right)$

15: **if** $\exists \widetilde{f} \in \widetilde{F} \mid \widetilde{f} \sim f'$ **then**

16: Set $f' = \widetilde{f}$

17: **else**

18: Set $unprocessed = unprocessed \cup \{f'\}$

19: Set $\widetilde{F} = \widetilde{F} \cup \{f'\}$

20: **if** $\left(f, r_k^{(j)}, f'\right) \notin \widetilde{\Delta}$ **then**

21: Set $\widetilde{\Delta} = \widetilde{\Delta} \cup \left\{\left(f, r_k^{(j)}, f'\right)\right\}$

23: Set $M\left(\left(f, r_k^{(j)}, f'\right)\right) = 1$

24: **else**

25: Set $M\left(\left(f, r_k^{(j)}, f'\right)\right) = M\left(\left(f, r_k^{(j)}, f'\right)\right) + 1$

26: **return** $\widetilde{\mathcal{S}} = \left(D, \widetilde{F}, R, \widetilde{\Delta}, M, \mathcal{C}, \mathcal{D}, f_{init}\right)$

The reduced state-space graph $\widetilde{\mathcal{S}}$ of the system illustrated in Figure 2 generated using Algorithm 2 is shown in Figure 4B. Compare this with the full state-space graph $\mathcal{S}$ generated using Algorithm 1, shown in Figure 4A.

The critical difference between Algorithm 1 and Algorithm 2 lies in line 15 of Algorithm 2. Line 15 tests whether a transition destination state $f'$ is equivalent to any representative system state $\widetilde{f} \in \widetilde{F}$ under the relation $\sim$. If it is, then $f'$ is set equal to $\widetilde{f}$ rather than adding $f'$ to the set of system states as is done in Algorithm 1. Methods for testing the equivalence of two states $f_1$ and $f_2$ are discussed in Appendix A2.

The reduced state-space graph $\widetilde{S}$ may be mapped to a CTMC using the general procedure described above of assigning rates to the transition labels using a weighting function; however, the multiplicity function $M$ must also be accounted for. In this case, the master equation associated with the resulting CTMC is derived by imposing an order on $\widetilde{F}$ using an indexing function $\widetilde{J}$ such that $\widetilde{J}:\widetilde{I} \rightarrow \widetilde{F}$, where $\widetilde{I} = \{1, 2, ..., |\widetilde{F}|\}$ is the index set, and is given by:

$$\frac{d\widetilde{P}}{dt} = \widetilde{A}\widetilde{P}, \tag{3}$$

where $\widetilde{P} = \left[\widetilde{P}_i\right]$ is a column vector defined such that $\widetilde{P}_i$, under the restriction that $\widetilde{P}_i > 0$ and $\sum_i \widetilde{P}_i = 1$, is the probability of the system occupying a state in equivalence class $\widetilde{J}(i)$ at time *t*, and $\widetilde{A} = \left[\widetilde{A}_{ij}\right]$ is a matrix of transition rates given by:

$$\widetilde{A}_{ij} = \begin{cases} M(\delta) \cdot W(r,t) & \text{for } \delta = \widetilde{J}(j) \xrightarrow{r} \widetilde{J}(i) \in \widetilde{\Delta}, \\ -\sum_{i \neq j} \widetilde{A}_{ij} & \text{for } i = j, \text{ and} \\ 0 & \text{otherwise,} \end{cases} \tag{4}$$

where $W(r,t)$ is the weight (transition rate) associated with label $r \in \bigcup_j R_j$ at time *t*. Since symmetries imply backward (as well as forward) bisimilarity of equivalent states under ~, the lumping of states in the resulting CTMC is exact such that the occupancy probability $P_i$ of any state $J(i) \in \widetilde{J}(j)$ is given by

$$P_i = \frac{P_j}{\left|\widetilde{J}(j)\right|},$$

where $\left|\widetilde{J}(j)\right|$ denotes the cardinality of equivalence class $\widetilde{J}(j)$.

***Invariant Manifold Reduction Technique***

An invariant manifold of an equation of the form of Eq. 1 is a vector $U = [U_i]$ parameterized by a variable *u* such that $P = U(u)$ is an exact solution of Eq. 1 when *u* satisfies a differential equation of form

$$\frac{du}{dt} = h(u,t) \tag{5}$$

of lower dimension than Eq. 1. Clearly, *U* is an invariant manifold of Eq. 1 if and only if the following holds[11]:

$$AU = \sum_j \frac{\partial U}{\partial u_j} h_j(u,t). \tag{6}$$

(For more discussion on some properties of invariant manifolds, see Ref. 11.)

It may be possible to achieve an exact reduction of Eq. 1 beyond what can be achieved through the symmetry reduction technique alone if an invariant manifold can be found. Although the existence of such an invariant manifold is an unresolved question in the general case, examples have been found in several specific cases[11,18]. Particularly relevant here is in the case of independence, i.e. when there are no interactions, neither functional nor synchronizing, between disjoint subsets of system components. More specifically, let $\delta \subseteq D$ be a subsystem with state-space graph given by:

$$\mathcal{S}_\delta = (\delta, F_\delta, L, \Delta_\delta, \mathcal{C}, \mathcal{D}, f_{init}).$$

Then, if *D* can be partitioned into a set $\{\delta_1, \delta_2, ..., \delta_n\}$ of subsystems with state spaces $\{F_{\delta_1}, F_{\delta_2}, ..., F_{\delta_n}\}$ such that all components in subsystem $\delta_i$ are independent from (i.e. have no functional or synchronizing interactions between) all components in subsystem $\delta_j$ for $j \neq i$, then an invariant manifold of the master equation associated with the CTMC generated on *D* exists, and is the product form solution given by:

$$P = U(P_1, P_2, ..., P_n) = \prod_{i=1}^{n} P_i \tag{7}$$

where each $P_i$ is a probability distribution associated with subsystem *i*. (Note that this invariant manifold holds for both the steady-state and transient solutions provided that initial transients have decayed.) It is straightforward to test for independence of subsystems using the HLM (see Appendix A3 for a procedure and associated algorithms).

The symmetry reduction and invariant manifold reduction techniques are not mutually exclusive. A general approach to combine the methods is to first identify independent subsystems, and to then apply symmetry reduction during the state-space generation for each subsystem using Algorithm 2. Applying this approach allows for potentially significant reductions in models of biochemical reaction systems *a priori*, as illustrated by the examples given below.

## IV. Examples

### *Example 1 – Single Ion Channel*

To illustrate the method as it is applied to the modeling of a biochemical process, consider a cluster of four identical $Na^+$ channels arranged in a ring configuration. Suppose that we first wish to develop a single channel model in which the individual channels are composed of one $\alpha$ subunit and two $\beta$ subunits, each of which may reside in one of the following possible conformations: inactive; permissive; or open. In addition, the $\beta$ subunits bind to a sodium ion and can also form dimers by binding to one another. We assume that $Na^+$ binds to the intracellular side of the channel.

We begin by modeling the $\beta$ subunit, which we divide into three functional domains: a $Na^+$ binding domain with two possible states (unbound and bound); a $\beta$ subunit binding domain with two possible states (unbound and bound); and a hinge domain with three possible states (inactive, permissive, open). Thus, we require three system components for each $\beta$ subunit. Because the $\alpha$ subunit does not have any binding domains, it may be modeled using a single component with three possible states (inactive, permissive, and open). The component state-space graph $\mathcal{C}$ for the channel may be formulated as shown in Figure 6 and Table 3. We assign the initial system state $f_{init}$ of the seven system components as (1,1,4,4,6,6,8), i.e. both $\beta$ subunits and the $\alpha$ subunit in the inactive state and both binding domains of the two $\beta$ subunits in the unbound state. According to $\mathcal{C}$, there are 10 possible component transitions for the $Na^+$ channel, four of which represent synchronizing transitions (i.e. for $j$ = 2, 3, 7, and 8), and five of which represent functional transitions (i.e. for $j$ = 2, 5, 6, 7, and 8).

The functional transitions require rules for their mapping to transition labels (rates), given in the form of dependency graphs. The dependency graphs are listed in Table 4. The first dependency graph $\mathcal{D}\left(r_1^{(2)}\right)$ encodes the rule that the two $\beta$ subunits must not be associated in order for the channel to open. $\mathcal{D}\left(r_1^{(5)}\right)$ and $\mathcal{D}\left(r_2^{(5)}\right)$ encode the rule that $Na^+$ binds to a $\beta$ subunit at rate $r_1^{(5)}$ if the channel is not open and at rate $r_2^{(5)}$ if the channel is open (the biophysical mechanism being that the local $Na^+$ concentration is elevated when the channel is open). $\mathcal{D}\left(r_1^{(6)}\right)$ and $\mathcal{D}\left(r_2^{(6)}\right)$ encode the rule that $Na^+$ dissociates with the rate $r_1^{(6)}$ if the $\beta$ subunit binding domain is unbound and with the rate $r_2^{(6)}$ if the $\beta$ subunits are in a dimer configuration (i.e. the affinity for $Na^+$ depends on whether the $\beta$ subunits are in a dimer configuration or not). $\mathcal{D}\left(r_1^{(7)}\right)$ encodes the rule that the $\beta$ subunits can only form dimers when the channel is closed. Finally, $\mathcal{D}\left(r_1^{(8)}\right)$, $\mathcal{D}\left(r_2^{(8)}\right)$, and $\mathcal{D}\left(r_3^{(8)}\right)$ encode the rule that the dissociation rate of the $\beta$ subunit dimer depends on whether $Na^{+}$ is bound to one, both, or none of the $\beta$ subunits.

Vernan, our MATLAB implementation of the method, can be used as a convenient platform for specifying and constructing the model described above (codes are available upon request). MATLAB codes for specifying and constructing the model using Vernan are given in the supplementary material. (In addition, a mat-file (Example1.mat) containing the model specifications is included in the Vernan package.) Vernan gives the option of generating the full state-space graph (using Algorithm 1) or the reduced state-space graph (using Algorithm 2). Generating the full state-space graph yields a model with 68 system states and 400 system transitions. (Recall that the full system state-space includes only those states that are reachable from the initial state $f_{init}$ which is why the full state-space size is much less than the product $3^2 \cdot 2^2 \cdot 2^2 \cdot 3 = 432$ of the number of states of each component.) On the other hand, generating the reduced state-space graph generates a model with 39 states and 178 transitions, a substantial reduction compared to the full state-space. Because the characteristic graph of this system is connected (i.e. there are dependencies between each component, and no two subsets of components are independent), it is not possible to achieve a further reduction of this system using the techniques described here.

The solution to the reduced model can be used to compute an exact solution to the full model using a differential equation of order 39, rather than 68 (or 432), which would be required without applying the reduction technique. To illustrate this, we assign random values in the interval (0,10) to 12 of the 14 rates in the model. We assign time-varying sinusoidal functions to the remaining two rates $r_1^{(5)}$ and $r_2^{(5)}$ such that $r_1^{(5)} \mapsto k_5 + A\left(1+\sin\left(\omega t\right)\right)$ and $r_2^{(5)} \mapsto k_6 + A\left(1+\sin\left(\omega t\right)\right)$, where $A$ is set to 5 second$^{-1}$, $\omega$ is set to $2\pi$ radians/second, and values for $k_5$ and $k_6$ are chosen randomly in the interval (0,10) with the constraint that $k_6$ be greater than $k_5$.

Next we numerically solve the master equations associated with the full and reduced systems (i.e. equations 1 and 3, respectively) using MATLAB's built-in solver ode15s using an initial probability of $P_i = 0$ for all $i$ such that $J(i) \in F \setminus f_{init}$ and $P_i = 1$ for the $i$ satisfying $J(i) = f_{init}$ in the full system, and of $\widetilde{P}_i = 0$ for all $i$ such that $\widetilde{J}(i) \in \widetilde{F} \setminus \left[f_{init}\right]$ and $\widetilde{P}_i = 1$ for the $i$ satisfying $\widetilde{J}(i) = \left[f_{init}\right]$ in the reduced system. (MATLAB codes are given in the supplementary material.) If we are interested in the open probability of the channel, we identify all states $f \in F$ and all states $\widetilde{f} \in \widetilde{F}$ such that the $\alpha$ subunit and two $\beta$ subunits are in the open conformation. There are four such $f$ in the full model, given by {(3,3,4,4,6,6,10), (3,3,4,5,6,6,10), (3,3,5,4,6,6,10), (3,3,5,5,6,6,10)}, and three such $\widetilde{f}$ in the reduced model, given by {(3,3,4,4,6,6,10), (3,3,4,5,6,6,10), (3,3,5,5,6,6,10)}. The macroscopic open probability is then simply the sum of the occupancy probabilities of these three (four) microscopic open states of the reduced (full) model. Plots of the time-evolution of the macroscopic open probability computed using the full and reduced models are shown in Figure 7A, demonstrating the equivalence of the two solutions.

Suppose now that we are interested in the occupancy probability of the microscopic state (3,3,5,4,6,6,10) and wish to compute this probability using the reduced model. Although this state is not included (i.e. is not a canonical representative) in the reduced system, we observe that this state is a member of the pattern of states represented by state (3,3,4,5,6,6,10) in the reduced system. The two states (3,3,4,5,6,6,10) and (3,3,5,4,6,6,10) together constitute the entire equivalence class represented by state (3,3,4,5,6,6,10). Thus, we can compute the occupancy probability of state (3,3,5,4,6,6,10) in the full system by simply dividing the occupancy probability of the pattern represented by (3,3,4,5,6,6,10) in the reduced system by two. The time-evolution of state (3,3,5,4,6,6,10) computed using the full and reduced system models, along with the time evolution of the state equivalence class represented by state (3,3,4,5,6,6,10), are plotted in Figure 7B.

*Example 2 – Ion channel complex with fine components*
In the next hypothetical example, the $Na^+$ channels modeled above are not physiologically observed alone, but rather as a complex of four such channels arranged in a ring configuration with each neighbor spaced equally apart. Whereas there were seven components in the individual channel model, there are now 28 ( $7 \times 4$ ) components in the complex model. We assign the initial state $f_{init}$ of the complex to be (1,1,1,1,1,1,1,1,4,4,4,4,4,4,4,4,6,6,6,6,6,6,6,6,8,8,8,8); that is, all four channels are in the inactive, unbound, dissociated state.

We assume that the four channels are close enough together that they are coupled through local $Na^+$ concentration elevations caused by the opening and closing of the individual $Na^+$ channels. Assuming that local $Na^+$ concentrations reach a steady-state on a time-scale much faster than $Na^+$ channels openings and closings, we can assume that this coupling is instantaneous. These are the same assumptions that are made in models of intracellular calcium release channel (ryanodine receptor) clusters[19]. This coupling depends on the spatial orientation of the channels, and because of our assumed orientation of the channels, there exist symmetries that can be exploited when constructing the $Na^+$ channel complex model.

The component state-space graph $\mathcal{C}$ for the complex model is the same as the individual channel model. However, there are now more rate constants associated with transition $Q_5$, i.e. the binding of a sodium ion to a $\beta$ subunit. Recall that in the single channel model, there were two possible binding rates depending on whether the channel was open or closed. In the channel complex model, there are 12 possible binding rates which depend on the open/closed configuration of all four channels in the complex. Consider the case where a sodium ion binds to a $\beta$ subunit of the channel labeled #1 in Figure 8. The 12 binding rates then depend on the open/closed configurations of all four channels as depicted in the diagram of Figure 8.

Another difference is that transition $Q_3$ is now a functional transition (recall that it was a constant transition in the single channel model). $Q_3$ is a synchronizing transition involving two $\beta$ subunits and one $\alpha$ subunit. The requirement that $Q_3$ be a functional transition in the complex model is a consequence of the fact that multiple channels may be in the open state simultaneously, allowing for $Q_3$ to occur between multiple combinations of components (whereas in the single channel model there was only one possible combination of components that could participate in $Q_3$). However, in our model, we assume that $Q_3$ can only occur between subunits of the same channel, and this requires a functional dependency to restrict the set of possible combinations of components that may be involved in the transition. Note that all synchronizing transitions in the complex model require a functional dependency for this reason, but all other synchronizing transitions were already functional transitions in the single channel model. Thus, it might have been easier to overlook this requirement for $Q_3$ compared to the others.

Since the system domain $D$ for the complex model is different from to the single channel model, new rules, along with a new set of dependency graphs $\mathcal{D}$, must be specified. Because the dependency graphs are rather complicated, we do not discuss them in detail here. However, a table listing the graphs can be found in the supplementary material. In addition, a .mat-file containing all specifications of the ion channel complex model, including the dependency graphs, as well as Vernan outputs, is included in the Vernan toolbox package (Example2.mat). Generating the reduced-state space graph using the Vernan tool generates a model with 304,590 system states and 5,414,760 system transitions, representing an over 70–fold reduction in state-space when compared to the expected state-space size ($68^4 = 21{,}381{,}376$) of the full model. It is not possible to achieve a larger reduction, as the characteristic graph of the model is connected. (For this example, it is not possible to generate the full model on a 32-bit system because the sparse transition matrix is too large to store in the available address space.)

*Example 3 – Ion channel complex with course components*

The large number of components in the ion channel complex model can make model specification a rather tedious task. However, there is flexibility in defining system components, which may be made as large or small as desired. In the preceding examples, the components were essentially specified to be as fine as possible (for the most part, constituting elementary reactions). In the case of the ion channel complex, it would have been possible to construct the model using only four components, where each component represents an individual ion channel. By doing this, the dependency graphs would be much smaller and easier to specify. However, the component state-space graph would be more complex. In addition, certain symmetries or independence between subsets of components that may exist in models specified using finer components may not exist in models specified using coarser components. Keeping this in mind, we next demonstrate the use of coarser components, namely the entire single channel model, in specifying the four channel complex model.

First, we must redefine the component state-space graph $\mathcal{C}$ as follows: we let the set of component states $S$ now be the set of state equivalence class representatives $\widetilde{F}$ of the reduced single-channel model and the set of component transitions $Q$ be the set of system transitions $\widetilde{\Delta}$ in the reduced single-channel model. All transitions in the new component state-space graph are local transitions. However, all transitions representing a $Na^+$ binding event in the single channel model must be made to be functional transitions in the complex model. Next, the rules and associated dependency graphs need to be defined. One difficulty that arises when specifying the rules for the coarser components is that, although there are many more component transitions, each of these transitions involves only one of a much smaller number of elementary reactions so that the number of labels (rates) is much smaller than the number of component transitions. Also, since we are dealing with the reduced single-channel model, a multiplicity is associated with each transition which must be accounted for. To avoid all of these difficulties, this model composition process can be automated using the Vernan tool. In Vernan, a function (SystemToComponent.m) is provided which takes the system state-space graph as input, and produces a new component state-space graph $\mathcal{C}$, a set of rules, and a mapping of transitions to a set of labels as output.

Finally, we define the new dependency graphs in $\mathcal{D}$ for the rates associated with the transition $Q_5$. (A .mat-file, Example3.mat, containing the specifications of this model and Vernan outputs is included in the Vernan toolbox package). In this case, there are only six graphs that need to be defined, which are shown in Table 5. Although $Na^+$ can bind with twelve different rates depending on the open/closed configuration of the four channels in the complex (as illustrated in Figure 8), any given channel only depends on the other channels, and there are only six possible configurations that these other channels can be in. With the finer components, we also had to account for the dependencies between subunits of the same channel, but these dependencies are implicitly accounted for by using the courser components.

With the new set of model specifications tailored for the larger components, we generate the reduced-state space graph of the system using Vernan, generating a model with 304,590 system states and 5,414,760 system transitions as before. If we index the system states of the two models in the same way, we can verify that the transition matrices associated with the two models are identical (see supplementary material for MATLAB codes), validating the use of courser components for specifying the model. There was a slight improvement in efficiency in generating the model using courser components. Where it took ~7 hours (on a desktop with 4-core 3.73 GHz Pentium D processors and 3.25 GB RAM) to generate the model using the 28 component specification, it only took ~5 hours using the 4 component specification. This speed up can be attributed to the smaller characteristic graph size when using the larger components (which would improve the efficiency of the canonical graph labeling routine), as well as to the smaller number of rules that needed to be checked when computing the state-space graph.

*Example 4 – Mean-field coupled ion channels*

Although it is not possible to achieve a further exact reduction of the ion-channel complex model described above using the techniques described here, it is certainly possible to make simplifying assumptions that would allow a further approximate reduction. One such simplifying assumption that is commonly applied to calcium-release site models is the mean-field coupling approximation[19,20], in which the local calcium concentration in a cluster of channels is assumed to be uniform across all channels so that the concentration at any one channel depends only on the numbers of open/closed channels in the cluster and not any specific spatial configuration of open/closed channels. We can apply this same approximation to our $Na^+$ channel complex model (here, we apply it to our 28 component model; model specifications and Vernan outputs are included in the file Example4.mat). By doing this, the number of different rates at which $Na^+$ binds to the $\beta$ subunit changes from 12 in the spatially-coupled model, to five in the mean-field coupled model. The rules by which transition $Q_5$ maps to these five rates must be defined, requiring us to re-specify our dependency graphs. The dependency graphs for all other rates remain the same.

As we have seen when specifying the spatially-coupled 28 component model, defining the dependency graphs can be a tedious process when a large number of components are involved. Specifying dependency graphs of a mean-field coupled model can be even more daunting, requiring a hyperarc for every possible configuration of open/closed channels. Although this process may be automated by writing a simple script, Vernan provides the option of defining dependency graphs in a mean-field coupling format (versus the spatially-coupled format which has been used up to now). Using the mean-field coupling format, we only need to specify one or more classes of component states along with the number of components required to occupy each of the classes of states. Dependency graphs in mean-field coupling format for the five rate constants mapping to $Q_5$ are given in Table 6 for the 28 component model.

Generating the reduced state-space graph of the newly specified system yields a model with 111,930 system states and 1,897,480 system transitions, an approximately three-fold reduction compared to the spatially-coupled model. This number of states is simply the number of combinations (with repetition) of the four channels on the set of 39 possible channel states in the reduced single channel model, given by the binomial coefficient

$$|F| = \binom{39+4-1}{4}.$$

Although we refer to the mean-field coupled model as an approximation, we emphasize that we are referring to an approximation in the mechanism of coupling, not an approximation in the solution of the master equation. Using the system specification of the mean-field coupled model, the 119,930 state model generated using Algorithm 2 is an exact reduction of the 21,381,376 state model that would be generated using Algorithm 1 given sufficient time and memory capacity.

*Example 5 – Independent ion channels*

As a final example, we assume that each individual channel in the complex is spaced far enough apart that the channels are effectively uncoupled from one another. In this case, the characteristic graph of the system becomes disconnected, and since no synchronizing transitions between channels exist, each channel is independent. (Vernan gives the option of testing for these conditions.) Then the channel complex model can be decomposed into four equivalent subsystems, each subsystem being the single channel model. In this case, an exact solution to the 21,382,376 state ion channel complex model may be obtained directly from the solution to the 39 state reduced single channel model. That is, the single channel model solution is the parameter of the invariant manifold of the ion channel complex model, given by Equation 7.

## Conclusion

### *Summary*

The system description formalism presented here offers a convenient and intuitive approach for specifying complex state transition systems with many different types of interactions. The technique allows the transition rules of a system to be systematically defined, helping to minimize the potential for modeling errors that could occur using a more direct manual approach. In addition, rule definitions do not require learning complicated language syntax and semantics, as they are easily understood in terms of simple graph theory. The way in which rules are specified allows a natural and automated application of the symmetry reduction[14] and invariant manifold reduction[11] techniques to achieve potentially significant exact reductions of master equations associated with the models.

To illustrate how the method may be applied to the modeling of biochemical systems, we have presented a hypothetical $Na^+$ channel complex model involving several different types of components and many different interactions. (All models were constructed using Vernan, our MATLAB implementation of the method (codes are available upon request).)  Using the reduction techniques described here, we generated a reduced model with an over 70-fold reduction in state-space when compared to the full (unreduced) version of the model. In general, highly symmetric dependencies among system components (as with the mean-field coupling approximation), or independence between components, results in large state-space reductions.

### *Improvements and Future Directions*

Although the reduction techniques described here focused on those that can be done *a priori* using a HLM, further exact reduction may be possible using techniques that operate directly at the state-level model. This further reduction may be possible for two reasons: 1) symmetries or independence between components exist at the state-space level that are non-identifiable at the system-description level; and 2) states exist that are either forward or backward bisimilar, but not both. The first situation can be avoided by careful modeling at the system-description level (see discussion in Appendix A3 for discussion). The second situation on the other hand cannot be managed at the system-description level and requires *a posteriori* processing to account for. In this case, the algorithm of Derisavi et al.[16] may be applied to compute the coarsest possible aggregation of the generated state-space.

In addition, it is also possible to compliment the exact reduction strategies described here with one or more approximate reduction strategies to achieve additional reduction of a model. Potential avenues for further reduction using approximation strategies include truncation[7] and probabilistic evaluation[8,9] techniques and perturbation methods such as time-scale separation[21,22], aggregation of nearly lumpable states[10], and decomposition of nearly independent subsystems. Probabilistic evaluation techniques such as the finite state projection algorithm of Munsky and Khammash[9] are particularly attractive as they offer the potential to automatically truncate the state-space of a very large, possibly infinite, state-space to a much smaller projection space of size just large enough to satisfy a user-specified tolerance in the total probability density error. (Although our system description method is restricted to systems with finite numbers of states and state transitions, it is possible to extend the specification method to allow specification of components with infinite state-spaces.) Algorithm 2 could be implemented directly in the finite state projection algorithm to expand the projection space on each iteration.

Finally, there is room for improvement in the efficiency of the reachability analyses described by Algorithms 1 and 2. In particular, Algorithm 2 would benefit from implementing one of the more efficient algorithms described by Junttila[14] for finding the canonical representative of a system state. (Using NAUTY[17] to find a canonical labeling of a characteristic graph is a bottleneck in computing the reachable state-space for systems with large characteristic graphs.) Also, computing the multiplicity for transitions with many possible combinations of reactants is costly using Algorithm 2, as it requires finding a canonical representative for the destination state of each combination. A better algorithm for computing the multiplicity function would therefore result in a significant speed up. Additionally, there is much room for improvement in the speed of our Vernan implementation of these algorithms by incorporating more efficient data structures and search methods.

**Acknowledgments**
This work is supported by the National Institute of Health under Grant P50-GM094503.

**Figure Legends**

**Table 1.** Component transitions $Q$ of the component state-space graph $\mathcal{C}$ for the example system illustrated in Figure 2. The mappings $Y_j$ and $Z_j$ are listed such that the first element in a set $X_j$ maps to the first elements listed in $Y_j$ and $Z_j$, the second element listed in a set $X_j$ maps to the second elements listed in $Y_j$ and $Z_j$, etc. Also given in the table are the sets of labels $R_j$ associated with each transition $j$.

**Table 2.** Dependency graphs for the rate constants of the example system illustrated in Figure 2. The labeling functions $L_y$ are listed such that the first element listed in a tail set $B_y$ maps to the first element listed in $L_y$, the second element listed in $B_y$ maps to the second element listed in $L_y$, etc.

**Table 3.** Component transitions $Q$ associated with the component state-space graph $\mathcal{C}$ shown in Figure 6. The mappings $Y_j$ and $Z_j$ are listed such that the first element in a set $X_j$ maps to the first elements listed in $Y_j$ and $Z_j$, the second element listed in a set $X_j$ maps to the second elements listed in $Y_j$ and $Z_j$, etc. Also given in the table are the sets of labels $R_j$ associated with each transition $j$.

**Table 4.** Dependency graphs for rate constants of the single channel model of Example 1. The labeling functions $L_y$ are listed such that the first element listed in a tail set $B_y$ maps to the first element listed in $L_y$, the second element listed in $B_y$ maps to the second element listed in $L_y$, etc.

**Table 5.** Dependency graphs for the rate constants associated with the binding of $Na^+$ to a $\beta$ subunit in the four component $Na^+$ sodium channel complex model. Each graph is associated with more than one rate label (due to the fact that multiple component transitions in the four component complex model correspond to a $Na^+$ binding event), which is why rate label superscripts are given as asterisks.

**Table 6.** Dependency graphs for the model of Example 4, given in Vernan's mean-field coupling format. In mean-field coupling format, a dependency $y$ is given as a class of states $\rho_y$, along with the number of components $N_y$ required to occupy the class of states in order to satisfy the dependency. Note that the dependencies labeled $y = 2$ in the five graphs are redundant and are not required to completely specify the model of Example 4. We include them here for completeness.

**Figure 1.** Example LTS. States in $F$ are given as circles and transitions in $\Delta$ are given as labeled arrows.

**Figure 2.** Example biochemical reaction system. *A*: An illustration of a small reaction volume containing two molecules of type A and one of type B. Molecules A can bind to molecules B, but only when in a folded conformation. *B*: An illustration of the component state space graph $\mathcal{C}$ associated with the reaction system depicted in panel *A*. Circles represent component states *S* and arrows represent component transitions *Q* which are labeled with a multiplicity. The system domain *D* and an initial system state $f_{init}$ are also shown.

**Figure 3.** Dependency graphs for the rate constants of the example system illustrated in Figure 2.

**Figure 4.** State-space graphs associated with the system illustrated in Figure 2. *A*: Full state-space graph $\mathcal{S}$. Circles represent the system states *F* and arrows represent the system transitions Δ. (Double arrows are used for neatness, and represent transitions in both the forward and reverse directions. The label for a transition of a given direction is the one nearest the arrowhead corresponding to that direction.) *B*: Reduced state-space graph $\widetilde{\mathcal{S}}$. Circles represent system state equivalence class representatives $\widetilde{F}$ and arrows represent the transitions $\widetilde{\Delta}$ between equivalence classes. The coefficients of the rate labels are the multiplicities of the transitions.

**Figure 5.** Characteristic graph $\mathcal{G}$ for the example system illustrated in Figure 2.

**Figure 6.** Component state-space graph $\mathcal{C}$ for $Na^+$ channel model. Circles represent component states *S* and arrows represent component transitions *Q* which are labeled with a multiplicity. The system domain *D* and an initial system state $f_{init}$ are also shown.

**Figure 7.** Transient solutions to the full and reduced versions of the single channel model. Solutions are for the system characterized by the set (ordered as listed in Table 3) of randomly chosen rate constants {4.5054, 0.8382, 2.2897, 9.1333, 0.0463+5*(1+sin(2*pi*t)), 7.7595+5*(1+sin(2*pi*t)), 1.5237, 8.2581, 5.3834, 9.9613, 0.7817, 4.4267, 1.0665, 9.6189}. *A*: Time-evolution of the open probability of the single channel model computed numerically from the full (dotted line) and reduced (solid line) versions of the model. *B:* Time-evolution of the occupancy probability of state $f = (3,3,5,4,6,6,10)$ computed numerically from the full (dotted line) and reduced (solid line) versions of the model, along with the time-evolution of the occupancy probability of the state pattern $[f]$, computed from the reduced model (dashed line).

**Figure 8.** The set of possible rate labels for the binding of $Na^+$ to the $\beta$ subunit of the channel labeled #1 in the $Na^+$ channel complex, along with the configurations of open and closed channels associated with each rate constant. Each channel is represented by a circle in the figure, and all channels are labeled as shown for $r_1^{(5)}$. Unfilled circles represent a channel in a closed state whereas filled circles represent a channel in an open state.

**Table 1**

$Q =$

| $j$ | $X_j$ | $Y_j$ | $Z_j$ | $R_j$ |
|---|---|---|---|---|
| 1 | {1} | (2) | (1) | $\{r_1^{(1)}\}$ |
| 2 | {2} | (1) | (1) | $\{r_1^{(2)}\}$ |
| 3 | {3,5} | (4,6) | (1,1) | $\{r_1^{(3)}\}$ |
| 4 | {4,6} | (3,5) | (1,1) | $\{r_1^{(4)}\}$ |

**Table 2**

| | $y$ | $A_y$ | $B_y$ | $L_y$ |
|---|---|---|---|---|
| $\mathcal{D}\left(r_1^{(2)}\right)$ | 1 | {1} | {3} | ({3}) |
| | 2 | {2} | {4} | ({3}) |
| $\mathcal{D}\left(r_1^{(3)}\right)$ | 1 | {3,5} | {1} | ({2}) |
| | 2 | {4,5} | {2} | ({2}) |

**Table 3**

$Q =$

| $j$ | $X_j$ | $Y_j$ | $Z_j$ | $R_j$ |
|---|---|---|---|---|
| 1 | {1} | (2) | (1) | $\{r_1^{(1)}\}$ |
| 2 | {2,9} | (3,10) | (2,1) | $\{r_1^{(2)}\}$ |
| 3 | {3,10} | (2,9) | (2,1) | $\{r_1^{(3)}\}$ |
| 4 | {2} | (1) | (1) | $\{r_1^{(4)}\}$ |
| 5 | {4} | (5) | (1) | $\{r_1^{(5)}, r_2^{(5)}\}$ |
| 6 | {5} | (4) | (1) | $\{r_1^{(6)}, r_2^{(6)}\}$ |
| 7 | {6} | (7) | (2) | $\{r_1^{(7)}\}$ |
| 8 | {7} | (6) | (2) | $\{r_1^{(8)}, r_2^{(8)}, r_3^{(8)}\}$ |
| 9 | {8} | (9) | (1) | $\{r_1^{(9)}\}$ |
| 10 | {9} | (8) | (1) | $\{r_1^{(10)}\}$ |

**Table 4**

| | $y$ | $A_y$ | $B_y$ | $L_y$ |
|---|---|---|---|---|
| $\mathcal{D}\left(r_1^{(2)}\right)$ | 1 | {1,2,7} | {5,6} | ({6},{6}) |
| $\mathcal{D}\left(r_1^{(5)}\right)$ | 1 | {3} | {1,2,7} | ({1,2},{1,2},{8,9}) |
| | 2 | {4} | {1,2,7} | ({1,2},{1,2},{8,9}) |
| $\mathcal{D}\left(r_2^{(5)}\right)$ | 1 | {3} | {1,2,7} | ({3},{3},{10}) |
| | 2 | {4} | {1,2,7} | ({3},{3},{10}) |
| $\mathcal{D}\left(r_1^{(6)}\right)$ | 1 | {3} | {5,6} | ({6},{6}) |
| | 2 | {4} | {5,6} | ({6},{6}) |
| $\mathcal{D}\left(r_2^{(6)}\right)$ | 1 | {3} | {5,6} | ({7},{7}) |
| | 2 | {4} | {5,6} | ({7},{7}) |
| $\mathcal{D}\left(r_1^{(7)}\right)$ | 1 | {5,6} | {1,2,7} | ({1,2},{1,2},{8,9}) |
| $\mathcal{D}\left(r_1^{(8)}\right)$ | 1 | {5,6} | {3,4} | ({4},{4}) |
| $\mathcal{D}\left(r_2^{(8)}\right)$ | 1 | {5,6} | {3,4} | ({4},{5}) |
| | 2 | {5,6} | {3,4} | ({5},{4}) |
| $\mathcal{D}\left(r_3^{(8)}\right)$ | 1 | {5,6} | {3,4} | ({5},{5}) |

**Table 5**

| | $y$ | $A_y$ | $B_y$ | $L_y$ |
|---|---|---|---|---|
| $\mathcal{D}(r_1^*)$ | 1 | {1} | {2,3,4} | (C,C,C) |
| | 2 | {2} | {1,3,4} | (C,C,C) |
| | 3 | {3} | {1,2,4} | (C,C,C) |
| | 4 | {4} | {1,2,3} | (C,C,C) |
| $\mathcal{D}(r_2^*)$ | 1 | {1} | {2,3,4} | (O,C,C) |
| | 2 | {1} | {2,3,4} | (C,C,O) |
| | 3 | {2} | {1,3,4} | (O,C,C) |
| | 4 | {2} | {1,3,4} | (C,O,C) |
| | 5 | {3} | {1,2,4} | (C,O,C) |
| | 6 | {3} | {1,2,4} | (C,C,O) |
| | 7 | {4} | {1,2,3} | (O,C,C) |
| | 8 | {4} | {1,2,3} | (C,C,O) |
| $\mathcal{D}(r_3^*)$ | 1 | {1} | {2,3,4} | (O,C,O) |
| | 2 | {2} | {1,3,4} | (O,O,C) |
| | 3 | {3} | {1,2,4} | (C,O,O) |
| | 4 | {4} | {1,2,3} | (O,C,O) |
| $\mathcal{D}(r_4^*)$ | 1 | {1} | {2,3,4} | (C,O,C) |
| | 2 | {2} | {1,3,4} | (C,C,O) |
| | 3 | {3} | {1,2,4} | (O,C,C) |
| | 4 | {4} | {1,2,3} | (C,O,C) |
| $\mathcal{D}(r_5^*)$ | 1 | {1} | {2,3,4} | (O,O,C) |
| | 2 | {1} | {2,3,4} | (C,O,O) |
| | 3 | {2} | {1,3,4} | (O,C,O) |
| | 4 | {2} | {1,3,4} | (C,O,O) |
| | 5 | {3} | {1,2,4} | (O,O,C) |
| | 6 | {3} | {1,2,4} | (O,C,O) |
| | 7 | {4} | {1,2,3} | (O,O,C) |
| | 8 | {4} | {1,2,3} | (C,O,O) |
| $\mathcal{D}(r_6^*)$ | 1 | {1} | {2,3,4} | (O,O,O) |
| | 2 | {2} | {1,3,4} | (O,O,O) |
| | 3 | {3} | {1,2,4} | (O,O,O) |
| | 4 | {4} | {1,2,3} | (O,O,O) |

**Table 6**

| | $y$ | $\rho_y$ | $N_y$ |
|---|---|---|---|
| $\mathcal{D}\left(r_1^{(5)}\right)$ | 1 | {10} | 0 |
| | 2 | {8,9} | 8 |
| $\mathcal{D}\left(r_2^{(5)}\right)$ | 1 | {10} | 1 |
| | 2 | {8,9} | 6 |
| $\mathcal{D}\left(r_3^{(5)}\right)$ | 1 | {10} | 2 |
| | 2 | {8,9} | 4 |
| $\mathcal{D}\left(r_4^{(5)}\right)$ | 1 | {10} | 3 |
| | 2 | {8,9} | 2 |
| $\mathcal{D}\left(r_5^{(5)}\right)$ | 1 | {10} | 4 |
| | 2 | {8,9} | 0 |

**Figure 1**

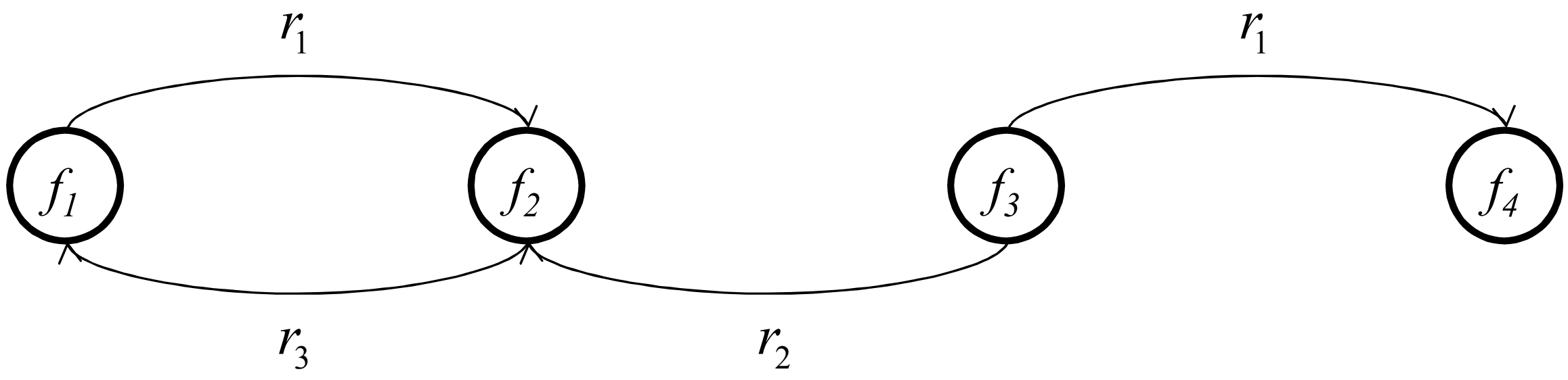

**Figure 2**

**A)**

**B)**

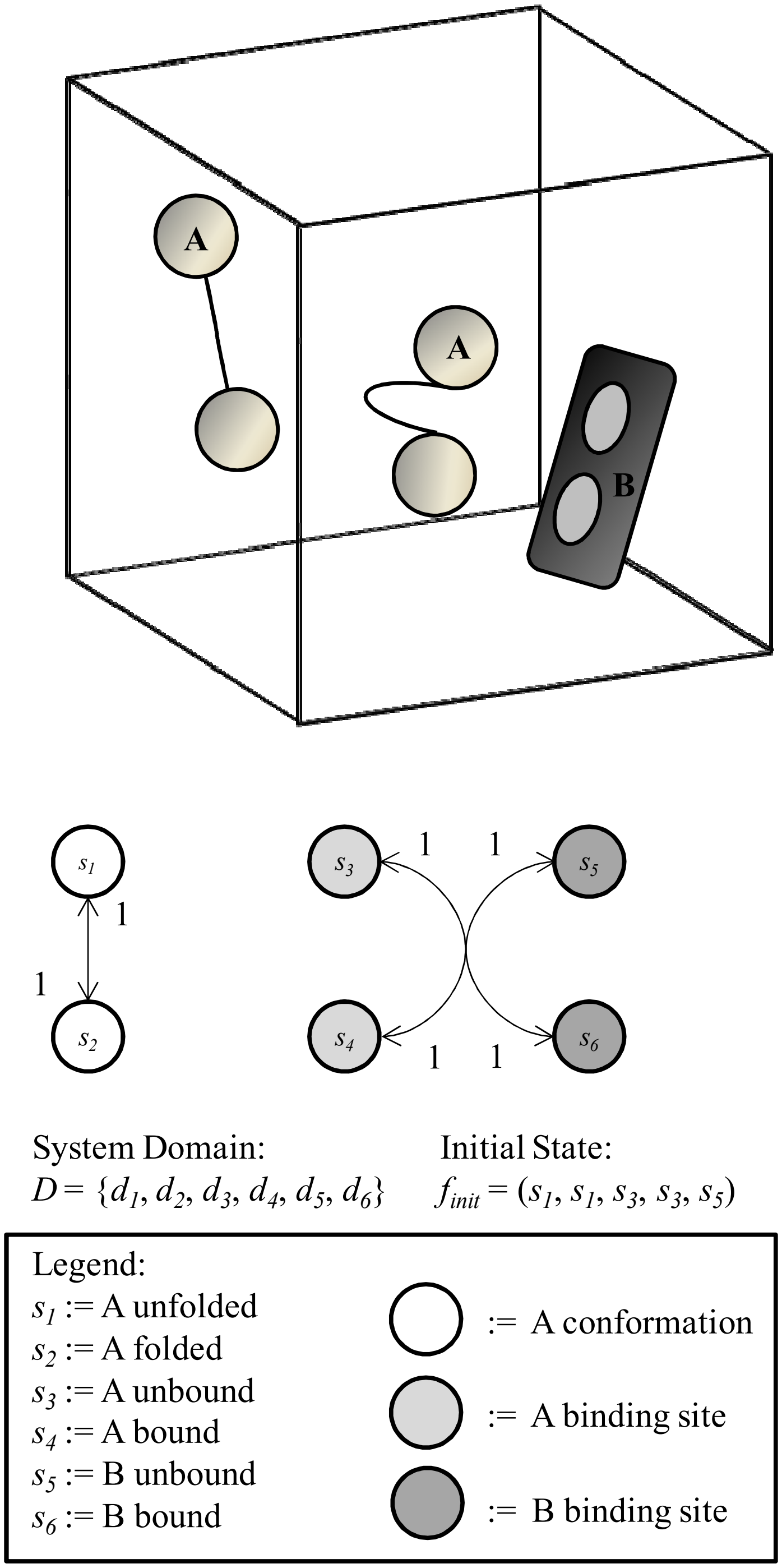
A
A
B
$s_1$
$s_2$
1
1
$s_3$
$s_4$
$s_5$
$s_6$
1 1
1 1
System Domain:
$D = \{d_1, d_2, d_3, d_4, d_5, d_6\}$
Initial State:
$f_{init} = (s_1, s_1, s_3, s_3, s_5)$
Legend:
$s_1$ := A unfolded
$s_2$ := A folded
$s_3$ := A unbound
$s_4$ := A bound
$s_5$ := B unbound
$s_6$ := B bound
:= A conformation
:= A binding site
:= B binding site

**Figure 3**

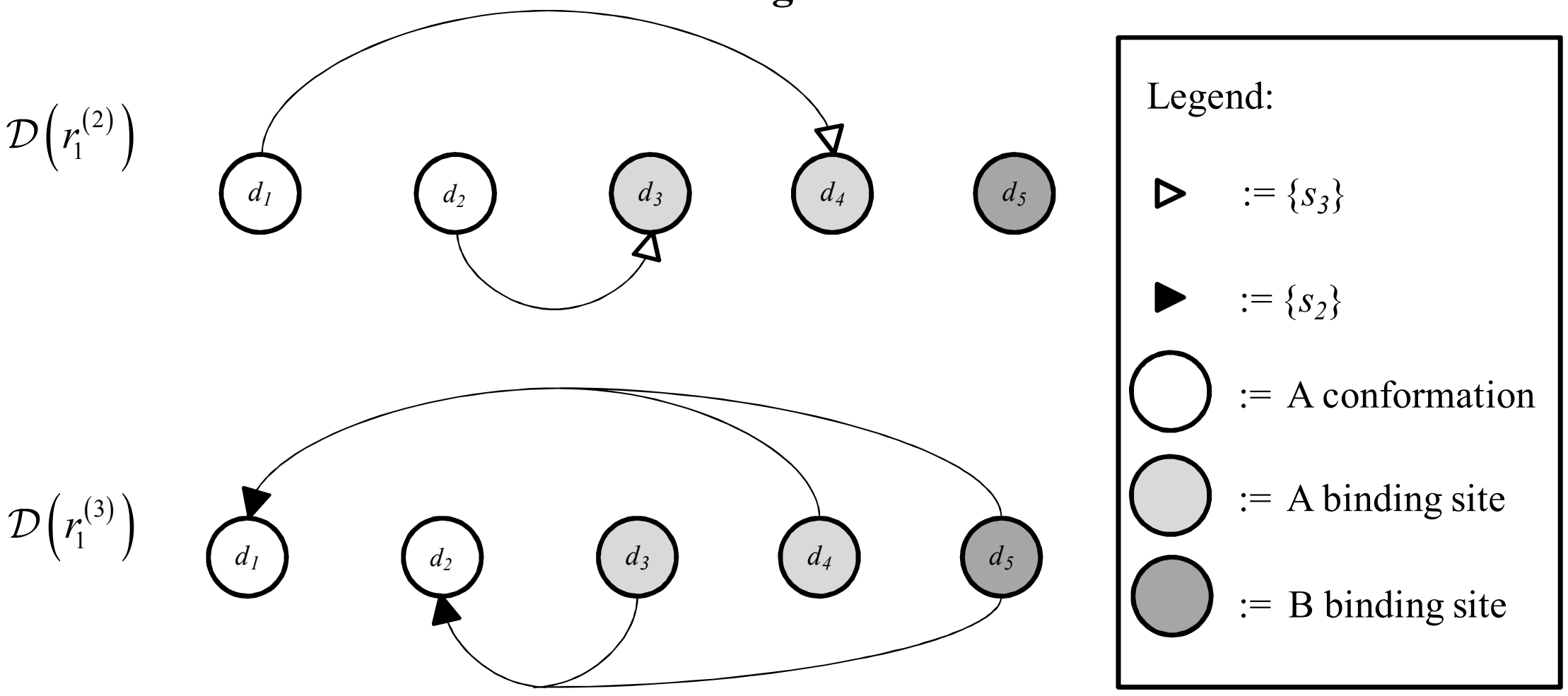

**Figure 4**

**A)**

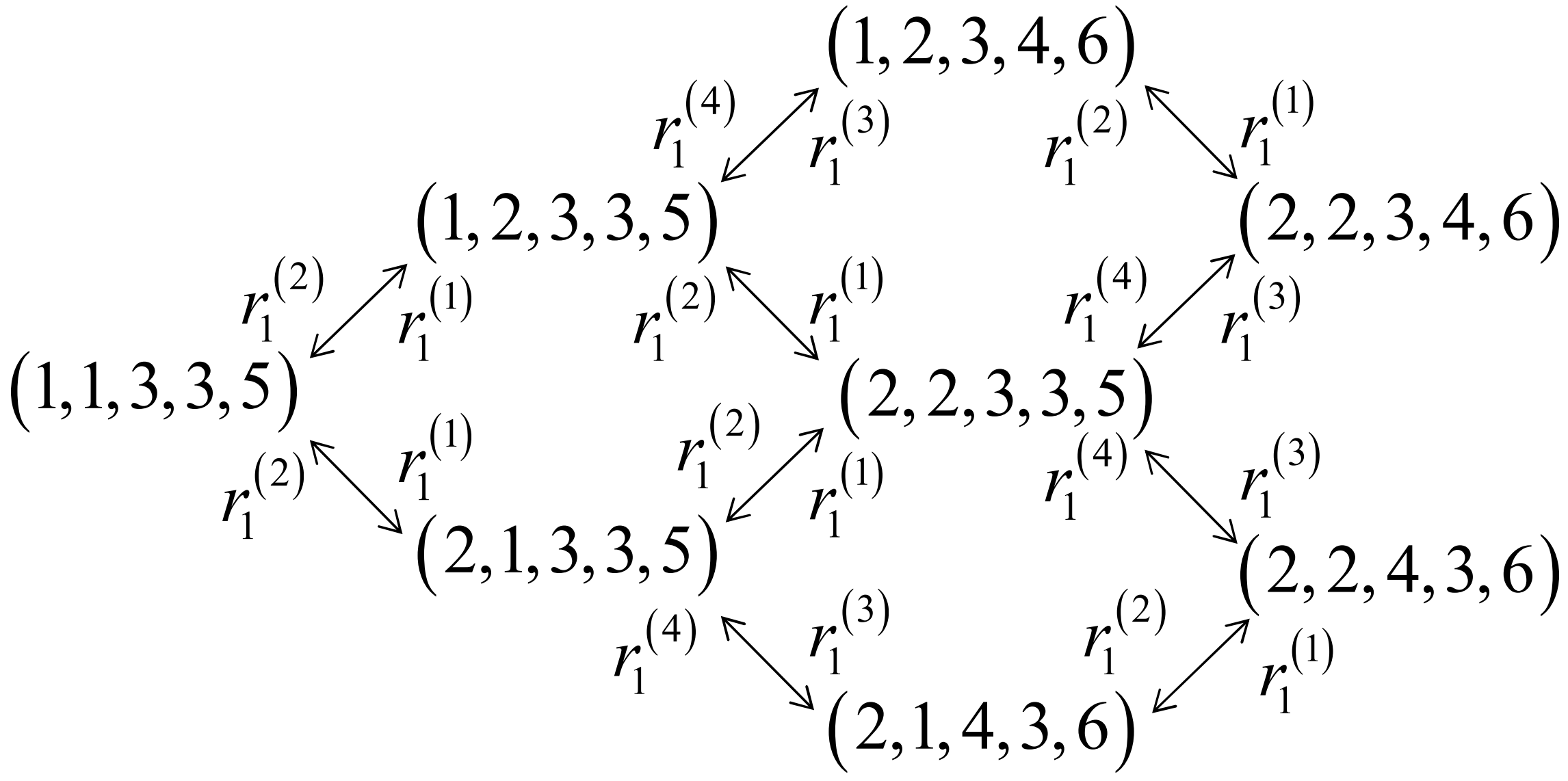
(1, 2, 3, 4, 6)
(1, 2, 3, 3, 5)
(2, 2, 3, 4, 6)
(1, 1, 3, 3, 5)
(2, 2, 3, 3, 5)
(2, 1, 3, 3, 5)
(2, 2, 4, 3, 6)
(2, 1, 4, 3, 6)
$r_1^{(1)}$ $r_1^{(2)}$ $r_1^{(3)}$ $r_1^{(4)}$

**B)**

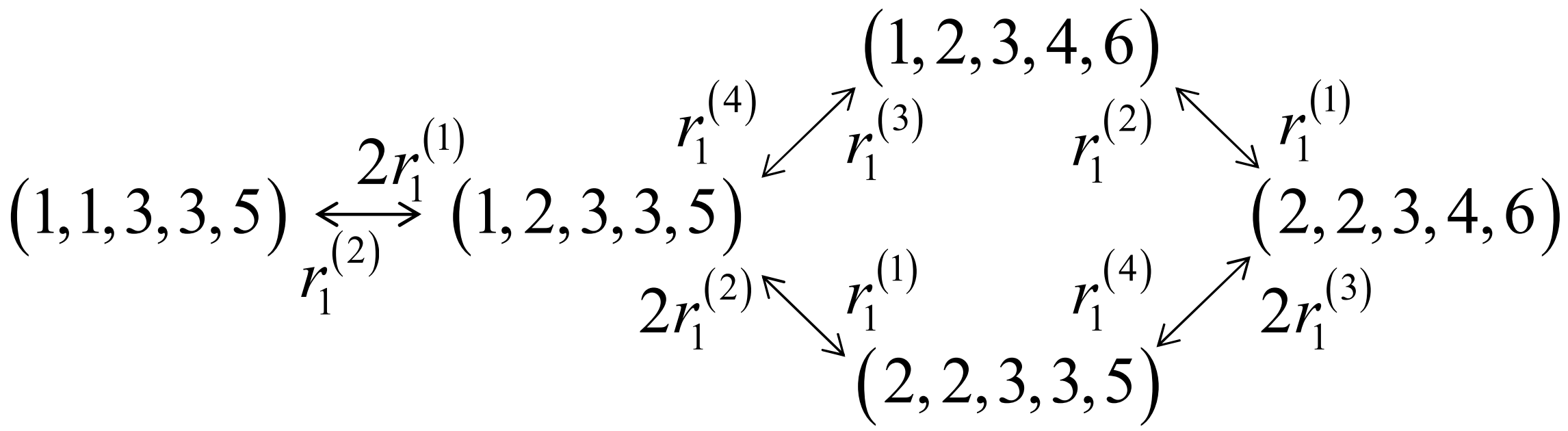
(1, 2, 3, 4, 6)
(1, 1, 3, 3, 5)
(1, 2, 3, 3, 5)
(2, 2, 3, 4, 6)
(2, 2, 3, 3, 5)
$2r_1^{(1)}$
$r_1^{(2)}$
$2r_1^{(2)}$
$2r_1^{(3)}$
$r_1^{(1)}$ $r_1^{(3)}$ $r_1^{(4)}$

**Figure 5**

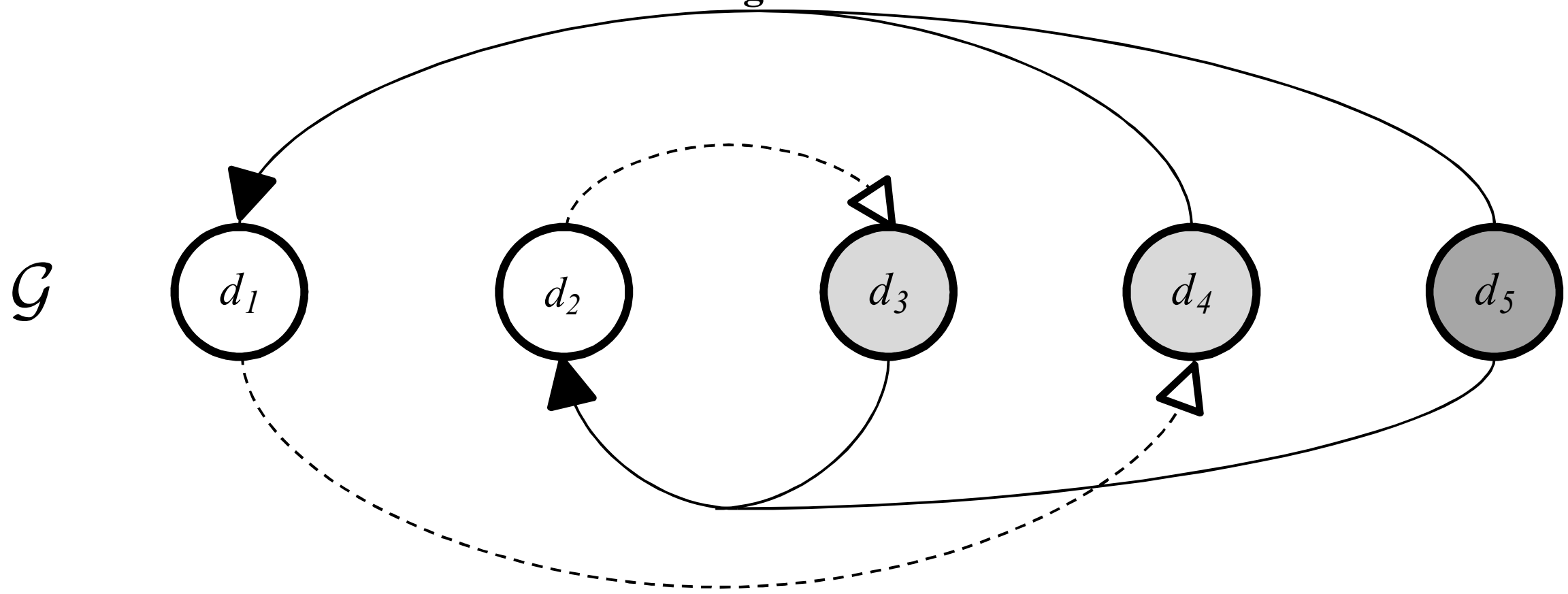

**Figure 6**

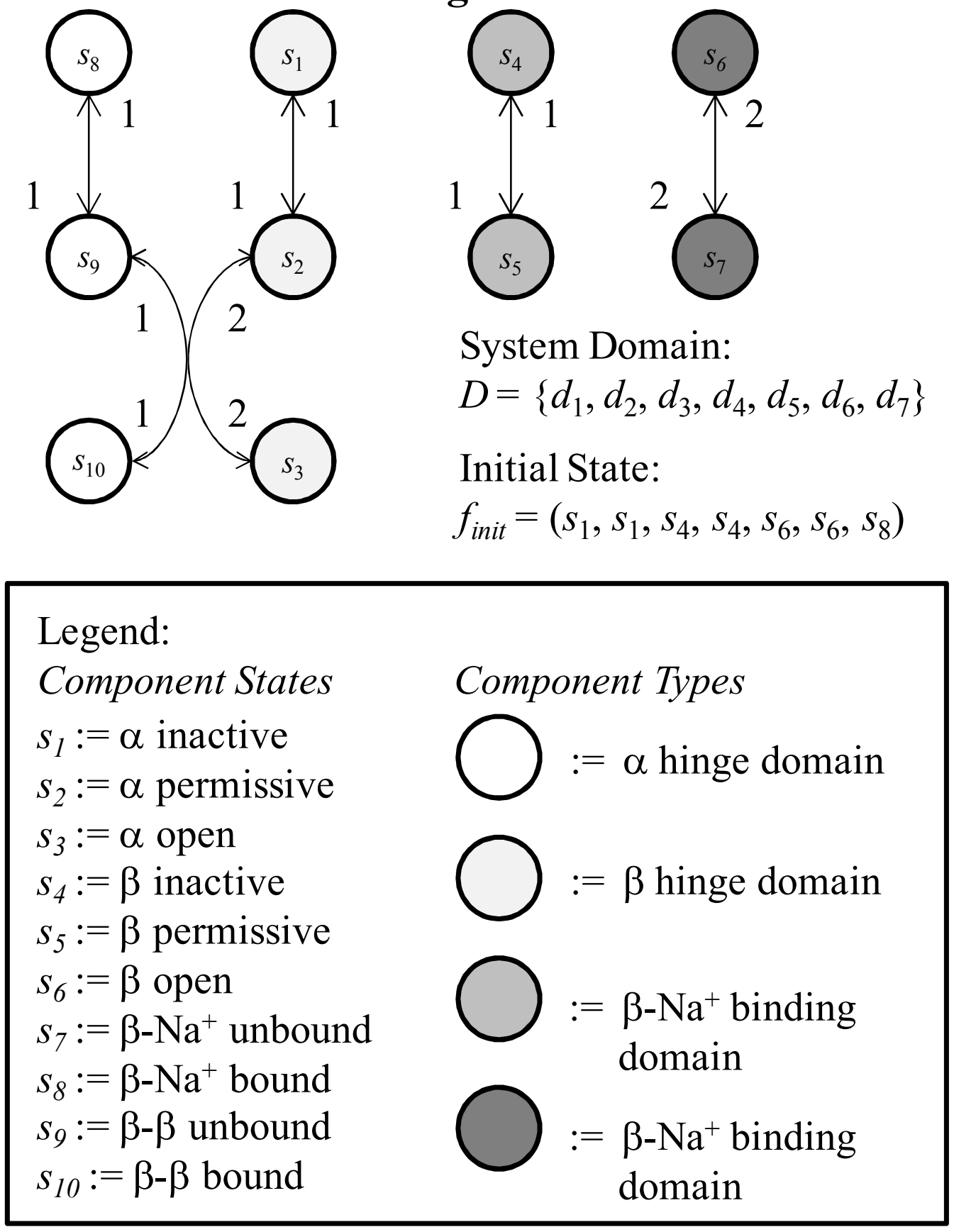

$s_8$ $s_1$ $s_4$ $s_6$
1 1 1 2
1 1 1 2
$s_9$ $s_2$ $s_5$ $s_7$
1 2
1 2
$s_{10}$ $s_3$
System Domain:
$D = \{d_1, d_2, d_3, d_4, d_5, d_6, d_7\}$
Initial State:
$f_{init} = (s_1, s_1, s_4, s_4, s_6, s_6, s_8)$
Legend:
*Component States*
$s_1$ := α inactive
$s_2$ := α permissive
$s_3$ := α open
$s_4$ := β inactive
$s_5$ := β permissive
$s_6$ := β open
$s_7$ := β-$Na^+$ unbound
$s_8$ := β-$Na^+$ bound
$s_9$ := β-β unbound
$s_{10}$ := β-β bound
*Component Types*
:= α hinge domain
:= β hinge domain
:= β-$Na^+$ binding domain
:= β-$Na^+$ binding domain

**Figure 7**

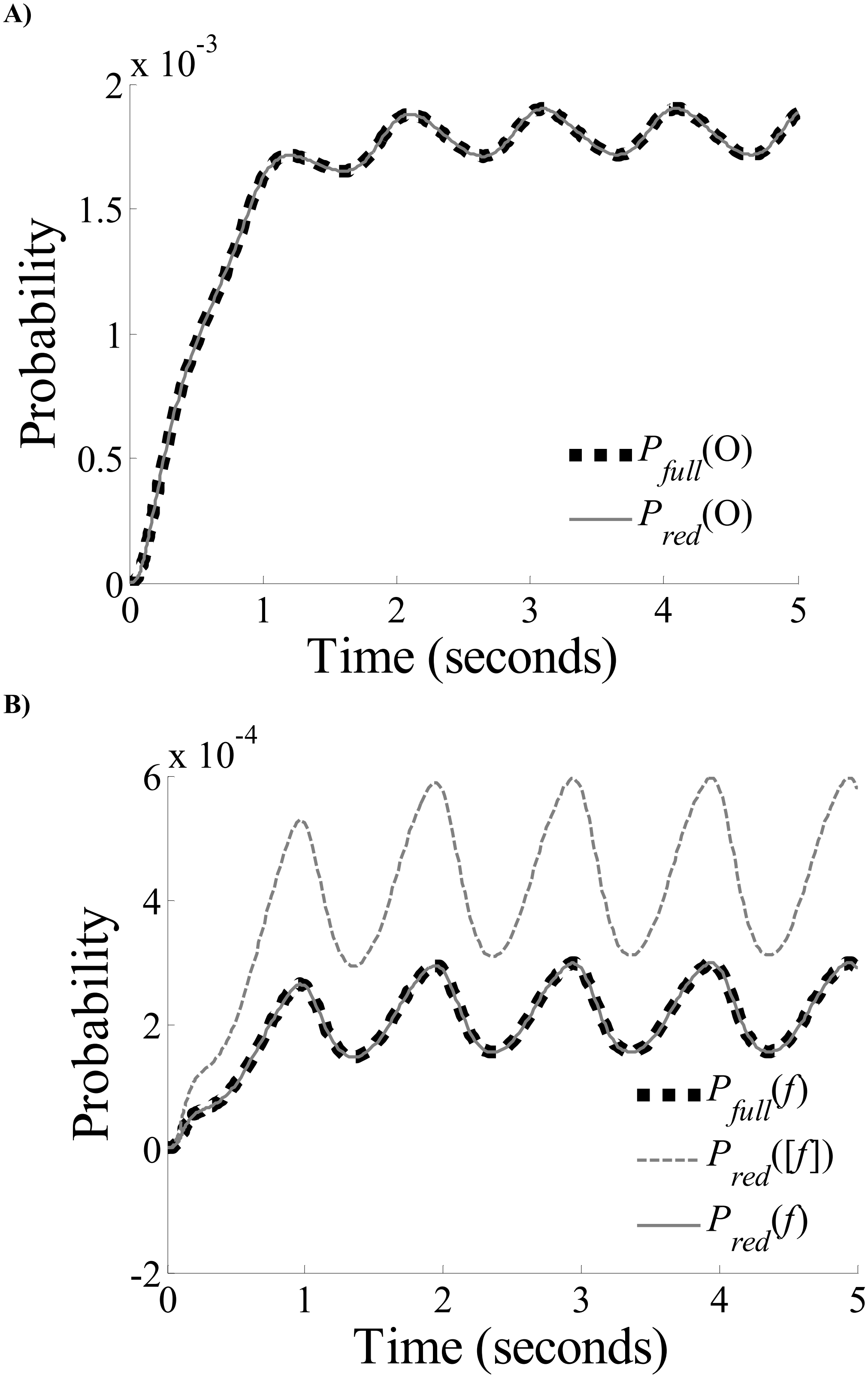

**Figure 8**

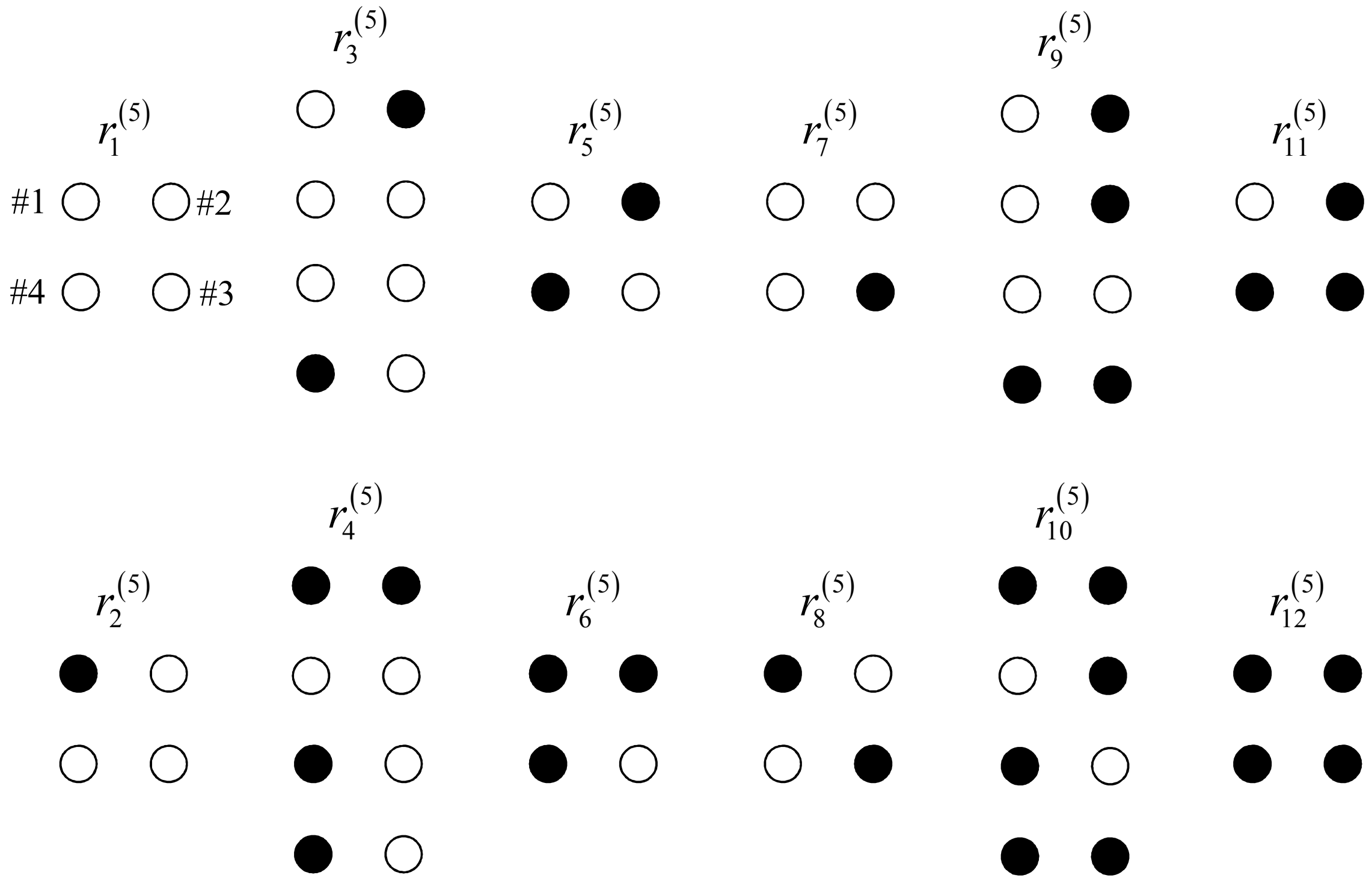

**A1.** Converting a directed arc-labeled hypergraph to an equivalent directed vertex-labeled simple graph.

Reaction dependencies, or rules, are specified in the form of directed arc-labeled hypergraphs. These graphs are used to construct a characteristic graph $\mathcal{G}$ for a system, as described in the text. In Vernan, automorphisms of $\mathcal{G}$ are identified using the NAUTY algorithm[17] which requires that graphs be given in simple vertex-labeled form. Converting $\mathcal{G}$ (a hypergraph) to an equivalent simple graph $\widetilde{\mathcal{G}}$ is done according to the procedure described below:

1. Generate a partition $\{U_1, U_2, ..., U_n\} \vdash U$ such that all hyperarcs belonging to a given $U_i$ have the same tail and head sets, and all hyperarcs belonging to $U_j$ for $j \neq i$ have tail and head sets different from those in $U_i$
2. For each $U_i$, define a labeling function $\mu_i$ assigning a multiset of labels to each head $d \in B_i$ such that:

$$\mu_i(d) = \uplus_l \left( L_l(d), r_k^{(j)} \right)$$

   where $L_l$ is the labeling function and $r_k^{(j)}$ is the rate label associated with hyperarc $E_l \in U_i$
3. For each $U_i$ containing hyperarcs with tail sets $A_i$ or head sets $B_i$ with cardinality greater than 1:
    a. Remove $U_i$ from $U$,
    b. Add a new vertex $h$ to $V$,
    c. Add arcs connecting all $d \in A_i$ to $h$,
    d. Add a new vertex $v_d$ for each $d \in B_i$, label each vertex according to $\mu_i(d)$, and add arcs connecting $h$ to $v_d$ and connecting $v_d$ to $d$.
4. For each $U_i$ containing hyperarcs with tail sets $A_i$ and head sets $B_i$ both with cardinality equal to 1:
    a. Remove $U_i$ from $U$,
    b. Add a new vertex labeled according $\mu_i(d)$, and add arcs connecting $A_i$ to the new vertex and connecting the new vertex to $B_i$.

Applying this conversion procedure to the characteristic graph $\mathcal{G}$ shown in Figure A1 panel A produces the equivalent simple graph $\widetilde{\mathcal{G}}$ shown in Figure A1 panel B. The automorphism group of $\widetilde{\mathcal{G}}$ restricted to the vertices labeled with numbers is the same as the automorphism group of $\mathcal{G}$.

**A2.** State equivalence testing and canonization method

There are several different ways of testing the equivalence of two states $f_1$ and $f_2$. A simple but naïve approach is to generate the full equivalence class $[f_1]$ of state $f_1$ by applying each element $g \in G$ to $f_1$, and to then test whether any function in $\left[f_1\right]$ is equal to $f_2$. This method works fine if $G$ is of low order but quickly becomes problematic for larger order $G$. For example, consider the case where all $p$ components of a system are identical symmetrically (that is, all permutations of the $p$ components are symmetries). Then, the group $G$ is the symmetric group of order $p!$, which can be very large for even relatively small $p$.

A better approach is to compute the canonical representatives of the states $f_1$ and $f_2$, and to then test whether the canonical representatives are the same. Several alternative methods for computing canonical representatives are presented in Ref. 14. Each of these methods exploit a structure known as the Schreier-Sims representation of a group, which is a compact representation of a group in the form of a base and strong generating set that allows linear time computations with groups. In our implementation, we apply the method described in section 4.2 of Ref. 14, which is based on the use of a graph canonizer to find the canonical version of a system state's characteristic graph.

The characteristic graph of a system state $f$ is defined as a graph $\mathcal{G}_f$ whose vertex set contains $D$ such that, for all system states $f_1$ and $f_2$, the following holds[14]:

1. if $g \in G$ maps $f_1$ to $f_2$, then there is an isomorphism $\gamma$ from $\mathcal{G}_{f_1}$ to $\mathcal{G}_{f_2}$ such that $\gamma$ restricted to $D$ equals $g$, and
2. if $\gamma$ is an isomorphism from $\mathcal{G}_{f_1}$ to $\mathcal{G}_{f_2}$, then $\gamma$ restricted to $D$ belongs to $G$ and maps $f_1$ to $f_2$.

Thus, the characteristic graphs $\mathcal{G}_{f_1}$ and $\mathcal{G}_{f_2}$ of $f_1$ and $f_2$ are isomorphic if and only if $f_1 \sim f_2$.

The bulk of the work for finding characteristic graphs of system states is done by constructing the system characteristic graph $\mathcal{G}$. The characteristic graph $\mathcal{G}_f$ of a system state $f$ is simply $\mathcal{G}$ with vertices corresponding to $D$ labeled according to $f\left(d\right)$ for all $d \in D$. Next, a black-box graph canonizer , such as the one provided in the NAUTY tool[17], is applied to find the canonical version of $\mathcal{G}_f$. A function $\mathcal{K}$ is a graph canonizer if, for all graphs $\mathcal{G}_1$ and $\mathcal{G}_2$, the following holds[14]:

1. $\mathcal{K}\left(\mathcal{G}_1\right)$ is isomorphic to $\mathcal{G}_1$, and
2. $\mathcal{K}\left(\mathcal{G}_1\right) = \mathcal{K}\left(\mathcal{G}_2\right)$ if and only if $\mathcal{G}_1$ and $\mathcal{G}_2$ are isomorphic to each other.

By applying the canonizer $\mathcal{K}$ to a system state characteristic graph $\mathcal{G}_f$, we obtain the canonical version $\mathcal{K}(\mathcal{G}_f)$ of the graph along with an isomorphism $\gamma$ carrying $\mathcal{G}_f$ to $\mathcal{K}(\mathcal{G}_f)$.

Given a permutation group represented in Schreier-Sims form, Junttila describes the concept of compatible permutations and gives an algorithm (Algorithm 4.1 of Ref. 14) for enumerating permutations in the group which are compatible with a place valuation *pval* (a function of the form $pval : D \to \mathbb{N}$) under a multiset selector (a function from a multiset to a subset of its underlying set of elements such that each element in the subset is associated with a non-zero multiplicity). Junttila proves that, if *pval* is an injective function, and the multiset selector used is a function multiset selector (a multiset selector with a singleton image set), then one and only one unique permutation is compatible with *pval*. Junttila further proves that, by letting *pval* be equal to the isomorphism $\gamma$ (restricted to *D*), and by finding the unique permutation $g \in G$ compatible with *pval* under a function multiset selector, then the canonical form of the system state *f* may be found by applying the inverse of *g* to *f* (Theorem 4.5 of Ref. 14). That is, $\mathcal{K}(f) = g^{-1}(f)$ is the canonical representative of system state *f*.

Using the canonical representative approach, Algorithm 2 may be improved. Specifically, in line 15, instead of testing whether each destination state $f'$ is isomorphic to any state in $\widetilde{F}$, the canonical version $\mathcal{K}(f')$ of $f'$ may be computed and tested for membership in $\widetilde{F}$.

**A3.** Testing for independence of system components

Independence of subsets of system components implies the absence of both functional and synchronizing interactions between the subsets of components. Determining whether there are any functional interactions between subsets of components can be done at the system description level by examining the characteristic graph $\mathcal{G}$ of the system. Similarly, determining whether there are any synchronizing interactions between component subsets can be done by examining the component graph $\mathcal{C}$ in conjunction with $f_{init}$. A sufficient condition for the existence of independent subsets of components is then given by the following:

1. the system characteristic graph $\mathcal{G}$ is disconnected, and
2. there exists a maximally connected subgraph of $\mathcal{G}$ such that none of its components are involved in a synchronizing transition with the components of any other disjoint maximally connected subgraph of $\mathcal{G}$.

Criterion number 1 is straightforward to test using a standard graph search algorithm such as the one given below:

**Algorithm 3.** An algorithm for computing the connectivity of a characteristic graph.

Set $unprocessed = V$
**while** $unprocessed \neq \varnothing$ **do**
  Select any $source \in unprocessed$
  Set $\pi(source) = source$
  Set $queue = source$
  **while** $queue \neq \varnothing$ **do**
    Select any $v \in queue$ and set $queue = queue \setminus \{v\}$
    Set $unprocessed = unprocessed \setminus \{v\}$
    **for all** $E_y \in \{E_y \in U \mid v \in A_y \cup B_y\}$ **do**
      **for all** $d \in A_y \cup B_y \cap unprocessed \setminus queue$ **do**
        Set $\pi(d) = source$
        Set $queue = queue \cup \{d\}$
**return** $\pi$

This algorithm generates a partition $\{V_1, V_2, \ldots V_n\} \vdash V$ of the vertices $V$ of $\mathcal{G}$ into $n$ maximal connected subsets such that the value of $\pi(d)$ is the same for all $d \in V_j$ and different for all $d \notin V_j$. If the partition is non-trivial (i.e. $\pi[V]$ is not singleton), then criterion number 1 is satisfied.

Criterion number 2 is more difficult to test, requiring that a reachability analysis be performed on $f_{init}$ similar to Algorithms 1 and 2. However, a stricter test can be done at the system description level, therefore not requiring a traversal of the reachability graph. This involves testing whether a synchronizing transition involving components of disjoint maximal connected subsets even exists (whereas criterion number 2 allows such a transition to exist but requires that components not participate in it). An algorithm for the test is given below:

**Algorithm 4.** An algorithm to merge disconnected subgraphs of $\mathcal{G}$ based on the existence of synchronizing transitions involving components of the subgraphs.

1: Set $unprocessed = V$
2: **while** $unprocessed \neq \varnothing$ **do**
3: Select any $v \in unprocessed$ and set $unprocessed = unprocessed \setminus \{v\}$
4: **for all** $Q_j \in \left\{ Q_j \in Q \mid \sum_{s \in X_j} Z_j(s) > 1 \wedge \mathcal{T}\left(f_{init}(v)\right) = \mathcal{T}(s) \exists s \in X_j \right\}$ **do**
5: Select any $s \in \left\{ s \in X_j \mid \mathcal{T}\left(f_{init}(v)\right) = \mathcal{T}(s) \right\}$
6: **for all** $d \in unprocessed$ **do**
7: **if** $\pi(d) \neq \pi(v) \wedge \mathcal{T}\left(f_{init}(d)\right) = \mathcal{T}(s') \exists s' \in \left(X_j, Z_j\right) \setminus \{\mathrm{s}\}$ **then**
8: **for all** $d' \in \left\{ d' \in V \mid \pi(d') = \pi(d) \right\}$ **do**
9: Set $\pi(d') = \pi(v)$
10: **if** $\pi[V] \setminus \pi(v) = \varnothing$ **then goto** 11
11: **return** $\pi$

In line 7, $\left(X_j, Z_j\right) \setminus \{s\}$ is the multiset difference rather than the usual set difference.

The above is a union-find algorithm that merges subsets of the partition $\{V_1, V_2, \ldots V_n\} \vdash V$ (given by $\pi$) together when synchronizing transitions exist between subsets. If the updated partition is non-trivial ($\pi[V]$ is non-singleton), then both criteria of the independence condition are satisfied.

The test given by the above algorithm should be sufficient in most cases. Careful modeling, especially of $f_{init}$ and of $Q$, can ensure that the stronger test need never be performed. (Note that criterion number 2 can also be verified directly *a posteriori* by examining the reachability graph $\widetilde{\mathcal{S}}$ after it is generated. In general however, it is preferred that the full $\widetilde{\mathcal{S}}$ is never generated when a reduced version exists.)

When specifying a model at the system description level, we account for all possible interactions between system components. However, as already mentioned in the case of synchronizing transitions, there will be no interaction in the model unless the transition actually occurs in the reachability graph. This is also the case for functional transitions.

That is, a functional interaction does not exist in a model unless its associated functional transition actually occurs in the reachability graph.

This subtle but important point has implications regarding both criterion number 1 of the independence condition above as well as in the model symmetry. If we are accounting for a functional interaction in our system characteristic graph, and that interaction does not actually exist in the reachability graph, then it is possible that there are symmetries in the model that we are not accounting for. While it is possible to verify the existence of interactions *a posteriori* by examining the generated reachability graph $\widetilde{\mathcal{S}}$, careful modeling at the system description level can render this step unnecessary. For example, if a particular transition in $Q$ requires that four components of type $y$ be in state $x$ to occur, and there are only three such components specified in $f_{init}$, then it is obvious that the transition never occurs and should therefore be removed from $Q$.

When the independence condition is satisfied, an invariant manifold of Eq. 1 exists, and the full system can be decomposed into a number of unique orthogonal subsystems by first partitioning $D$ according to $\pi$. That is, $D$ is partitioned into $\{D_1, D_2, ...D_n\}$ such that the value of $\pi(d)$ is the same for all $d \in D_j$ and different for all $d \notin D_j$. Next, each subsystem can be characterized as unique or non-unique by testing the following:

1. whether any other subsystem contains the same number and type of components, and
2. whether its associated subsystem characteristic graph, defined as the induced subgraph of $\mathcal{G}$ with vertex set $D_j$, is isomorphic to any other subsystem's characteristic graph.

State-level models generated for each unique subsystem may be used to obtain an exact solution to the full model as described in the main text.

**Appendix Figure Legends**

**Figure A1.** Conversion of a hypergraph $\mathcal{G}$ to an equivalent vertex-colored simple graph $\widetilde{\mathcal{G}}$. $A$: A hypothetical characteristic graph $\mathcal{G}$. $B$: An equivalent simple graph $\widetilde{\mathcal{G}}$ obtained by applying the procedure described in Appendix A1. The additional vertices in the graph labeled with letters are the result of step 3b in the procedure whereas the smaller unlabeled vertices are the result of steps 3d and 4b. The smaller vertices are colored such that vertices assigned to the same multiset of labels by a function $\mu_i$ have the same color and those assigned to different multisets of labels have different colors.

**Figure A1**

**A**

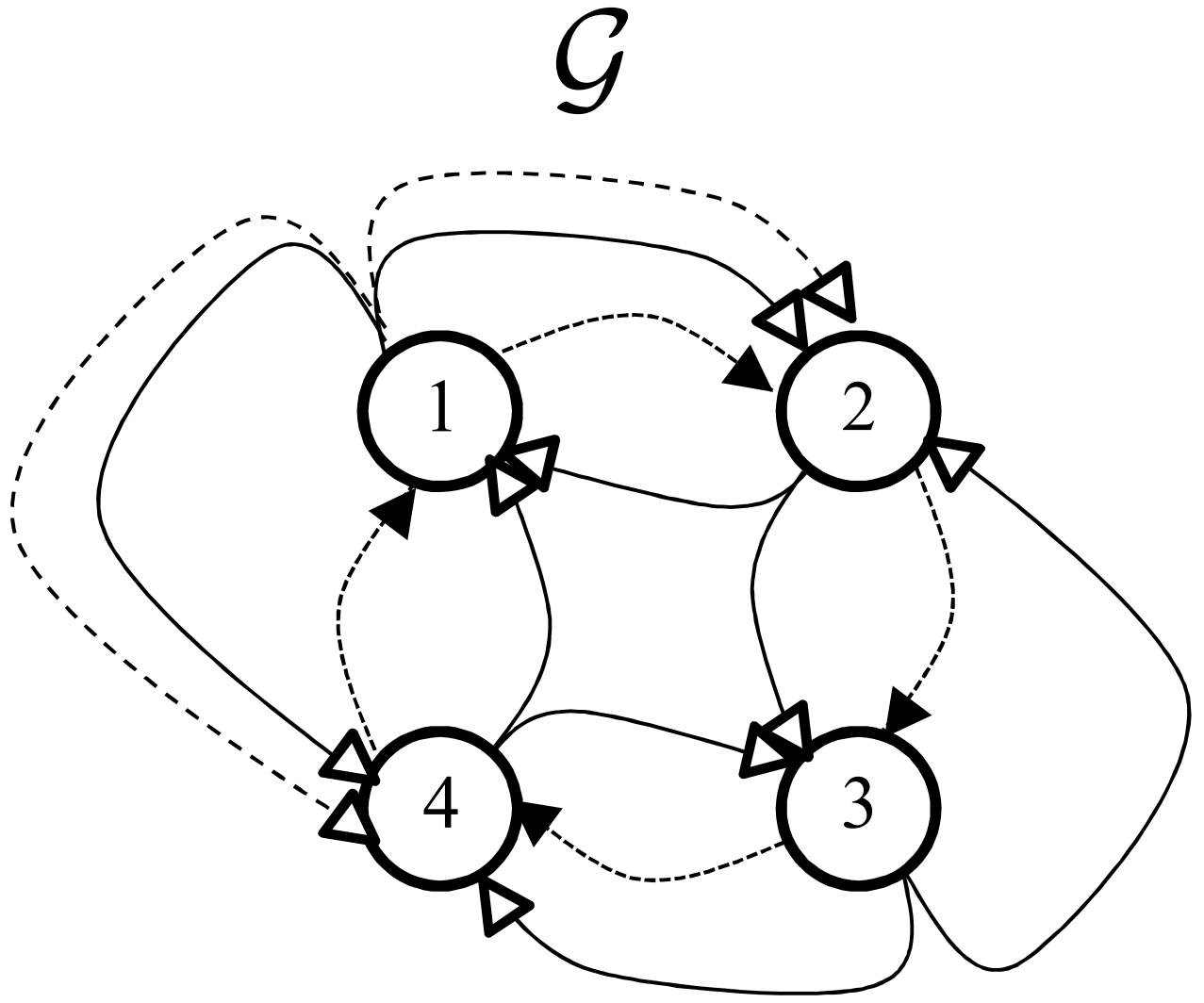

**B**

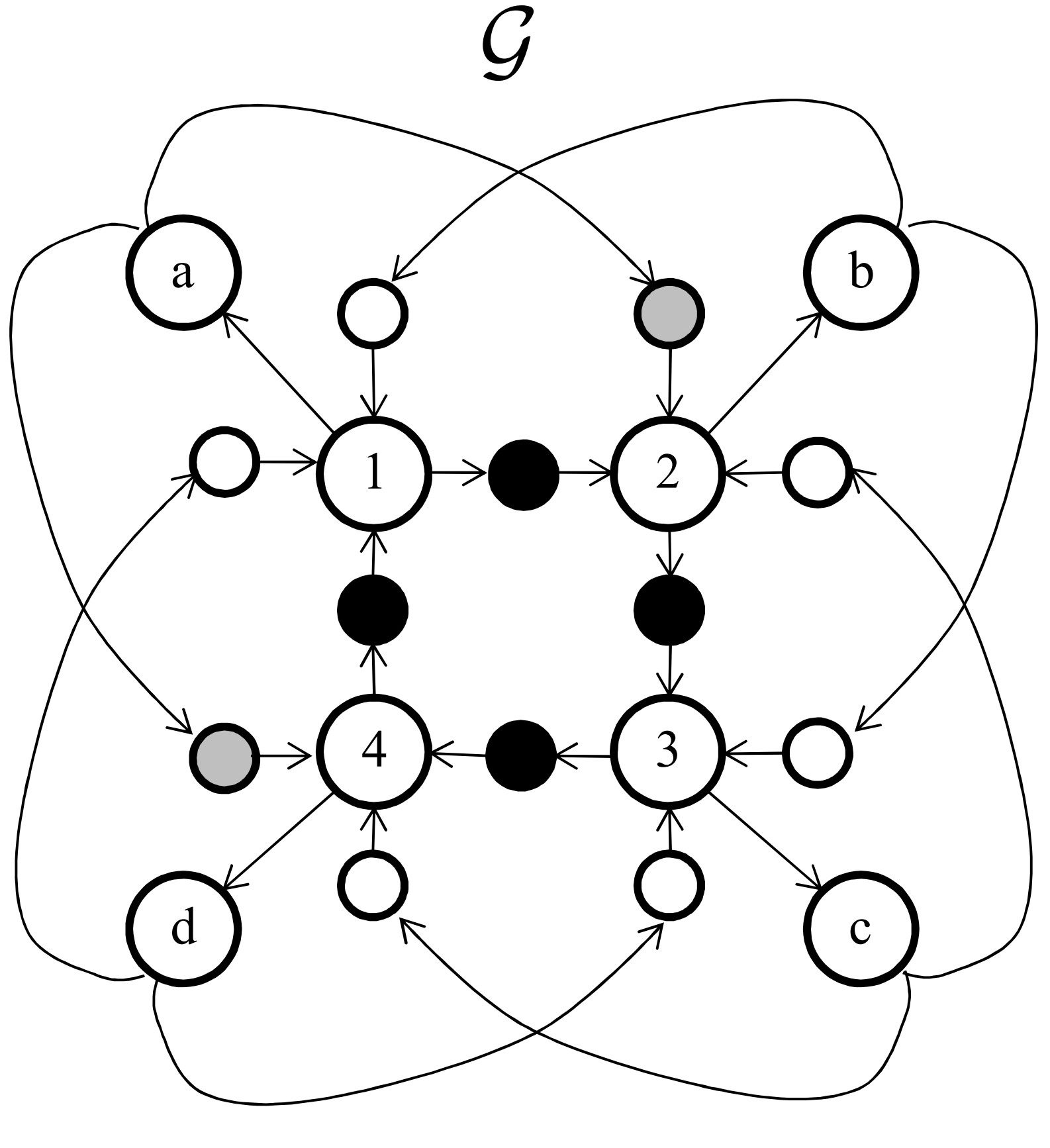

**S1.** MATLAB codes for specifying and generating a model of the system illustrated in Figure 2 using the Vernan toolbox.

```
%% Read Vernan documentation
help Vernan

%% Set-up inputs
n = 6;
Q = cell(4,3);

%   Xj                Yj                Zj
Q{1,1} = 1;       Q{1,2} = 2;       Q{1,3} = 1;
Q{2,1} = 2;       Q{2,2} = 1;       Q{2,3} = 1;
Q{3,1} = [3,5];   Q{3,2} = [4,6];   Q{3,3} = [1,1];
Q{4,1} = [4,6];   Q{4,2} = [3,5];   Q{4,3} = [1,1];

finit = [1,1,3,3,5];

rho = cell(1,2);
rho{1} = 2;  rho{2} = 3;

%Dependency graphs below are indexed according to the order the graphs are
%listed in Table 2
D = cell(2,1);
D{1} = cell(2,3);  D{2} = cell(2,3);

%     Ay                  By            Ry
D{1}{1,1} = 1;      D{1}{1,2} = 3;  D{1}{1,3} = 2;
D{1}{2,1} = 2;      D{1}{2,2} = 4;  D{1}{2,3} = 2;
%--------------------------------------------------
D{2}{1,1} = [3,5];  D{2}{1,2} = 1;  D{2}{1,3} = 1;
D{2}{2,1} = [4,5];  D{2}{2,2} = 2;  D{2}{2,3} = 1;

rules = zeros(14,4);
rules(1,:) = [1,1,0,0];  rules(2,:) = [2,2,1,0];
rules(3,:) = [3,3,2,0];  rules(4,:) = [4,4,0,0];

%% Generate state-space graphs
symred = 'off'; %generate full model (corresponds to Figure 2A)
[Ffull,Tfull,Wfull] = Vernan(n,Q,rules,D,rho,finit,symred);

symred = 'on'; %generate reduced model (corresponds to Figure 2B)
[Fred,Tred,Wred,M] = Vernan(n,Q,rules,D,rho,finit,symred);
```

**S2.** MATLAB codes for specifying, generating, and solving the full and reduced single $Na^+$ channel model of Example 1 using the Vernan toolbox.

```
%% Read Vernan documentation
help Vernan

%% Set-up inputs
n = 10;
Q = cell(10,3);

%   Xj               Yj                Zj
Q{1,1} = 1;       Q{1,2} = 2;       Q{1,3} = 1;
Q{2,1} = [2,9];   Q{2,2} = [3,10];  Q{2,3} = [2,1];
Q{3,1} = [3,10];  Q{3,2} = [2,9];   Q{3,3} = [2,1];
Q{4,1} = 2;       Q{4,2} = 1;       Q{4,3} = 1;
Q{5,1} = 4;       Q{5,2} = 5;       Q{5,3} = 1;
Q{6,1} = 5;       Q{6,2} = 4;       Q{6,3} = 1;
Q{7,1} = 6;       Q{7,2} = 7;       Q{7,3} = 2;
Q{8,1} = 7;       Q{8,2} = 6;       Q{8,3} = 2;
Q{9,1} = 8;       Q{9,2} = 9;       Q{9,3} = 1;
Q{10,1} = 9;      Q{10,2} = 8;      Q{10,3} = 1;

finit = [1,1,4,4,6,6,8];

rho = cell(1,8);
rho{1} = [1,2];  rho{2} = 3;  rho{3} = 4;  rho{4} = 5;
rho{5} = 6;      rho{6} = 7;  rho{7} = [8,9]; rho{8} = 10;

%Dependency graphs below are indexed according to the order the graphs are
%listed in Table 4
D = cell(9,1);
D{1} = cell(1,3);  D{2} = cell(2,3);  D{3} = cell(2,3);  D{4} = cell(2,3);
D{5} = cell(2,3);  D{6} = cell(1,3);  D{7} = cell(1,3);  D{8} = cell(2,3);
D{9} = cell(1,3);

%        Ay                   By                   Ry
D{1}{1,1} = [1,2,7];  D{1}{1,2} = [5,6];    D{1}{1,3} = [5,5];
%---------------------------------------------------------------
D{2}{1,1} = 3;        D{2}{1,2} = [1,2,7];  D{2}{1,3} = [1,1,7];
D{2}{2,1} = 4;        D{2}{2,2} = [1,2,7];  D{2}{2,3} = [1,1,7];
%---------------------------------------------------------------
D{3}{1,1} = 3;        D{3}{1,2} = [1,2,7];  D{3}{1,3} = [2,2,8];
D{3}{2,1} = 4;        D{3}{2,2} = [1,2,7];  D{3}{2,3} = [2,2,8];
%---------------------------------------------------------------
D{4}{1,1} = 3;        D{4}{1,2} = [5,6];    D{4}{1,3} = [5,5];
D{4}{2,1} = 4;        D{4}{2,2} = [5,6];    D{4}{2,3} = [5,5];
%---------------------------------------------------------------
D{5}{1,1} = 3;        D{5}{1,2} = [5,6];    D{5}{1,3} = [6,6];
D{5}{2,1} = 4;        D{5}{2,2} = [5,6];    D{5}{2,3} = [6,6];
%---------------------------------------------------------------
D{6}{1,1} = [5,6];    D{6}{1,2} = [1,2,7];  D{6}{1,3} = [1,1,7];
%---------------------------------------------------------------
D{7}{1,1} = [5,6];    D{7}{1,2} = [3,4];    D{7}{1,3} = [3,3];
%---------------------------------------------------------------
D{8}{1,1} = [5,6];    D{8}{1,2} = [3,4];    D{8}{1,3} = [3,4];
D{8}{2,1} = [5,6];    D{8}{2,2} = [3,4];    D{8}{2,3} = [4,3];
%---------------------------------------------------------------
D{9}{1,1} = [5,6];    D{9}{1,2} = [3,4];    D{9}{1,3} = [4,4];

rules = zeros(14,4);
rules(1,:) = [1,1,0,0];  rules(2,:) = [2,2,1,0];  rules(3,:) = [3,3,0,0];
rules(4,:) = [4,4,0,0];  rules(5,:) = [5,5,2,0];  rules(6,:) = [5,6,3,0];
rules(7,:) = [6,7,4,0];  rules(8,:) = [6,8,5,0];  rules(9,:) = [7,9,6,0];
rules(10,:) = [8,10,7,0];  rules(11,:) = [8,11,8,0];
rules(12,:) = [8,12,9,0];  rules(13,:) = [9,13,0,0];
rules(14,:) = [10,14,0,0];
```

```
%% Generate state-space graphs
symred = 'off'; %generate full model
[Ffull,Tfull,Wfull] = Vernan(n,Q,rules,D,rho,finit,symred);

symred = 'on'; %generate reduced model
[Fred,Tred,Wred,M] = Vernan(n,Q,rules,D,rho,finit,symred);

%% Solve ODEs
help SolveMasterEqtn_transient %read documentation

C = cell(14,1); % a set of 14 specific probability rate constants
for i = [1:4,7:14]
    C{i} = 10*rand(1);
end
k5 = 10*rand(1);
k6 = (10 - k5)*rand(1) + k5;
C{5} = @(t)k5 + 5*(1+sin(2*pi*t)); %C{5} and C{6} are made time-dependent
C{6} = @(t)k6 + 5*(1+sin(2*pi*t));

odesolver = @ode15s; % solve using MATLAB's built-in ode15s solver
tspan = [0,5]; % integrate over 5 seconds
options = odeset('RelTol',1e-12); % set tolerance

Pfull0 = zeros(length(Ffull),1);
Jfull = ismember(Ffull,finit,'rows'); % locates index of finit in Ffull
Pfull0(Jfull) = 1;

Pred0 = zeros(length(Fred),1);
Jred = ismember(Fred,finit,'rows'); % locates index of finit in Fred
Pred0(Jred) = 1;

[Pfull,tfull] = SolveMasterEqtn_transient(Ffull,Tfull,Wfull,1,C, ...
                odesolver,tspan,Pfull0,options); %solve full model

[Pred,tred] = SolveMasterEqtn_transient(Fred,Tred,Wred,M,C, ...
              odesolver,tspan,Pred0,options); %solve reduced model

%% Plot results
%plot time-evolution of open states
Jfull = ismember(Ffull(:,7),10); %to find open states, we locate the states
Jred = ismember(Fred(:,7),10);   %with component 7 in state 10
Pfull_open = sum(Pfull(Jfull,:),1);
Pred_open = sum(Pred(Jred,:),1);

figure(1); % corresponds to Figure 4A in text
plot(tfull,Pfull_open,'r',tred,Pred_open,':k');
xlabel('time (seconds)');
ylabel('Probability');

%plot time-evolution of state f = (3,3,5,4,6,6,10)
Jfull = ismember(Ffull,[3,3,5,4,6,6,10],'rows');
Jred = ismember(Fred,[3,3,4,5,6,6,10],'rows');
Pfull_f = Pfull(Jfull,:);
Pred_ftilde = Pred(Jred,:);
Pred_f = Pred(Jred,:)./2;

figure(2); % corresponds to Figure 4B in text
plot(tfull,Pfull_f,'r',tred,Pred_ftilde,'k',tred,Pred_f,':k');
xlabel('time (seconds)');
ylabel('Probability');
```

**S3.** Dependency graphs for the 28 component $Na^+$ channel complex model of Example 2.

| | $y$ | $A_y$ | $B_y$ | $R_y$ |
|---|---|---|---|---|
| $\mathcal{D}\left(r_1^{(2)}\right)$ | 1 | {1,2,25} | {17,18} | ({6},{6}) |
| | 2 | {3,4,26} | {19,20} | ({6},{6}) |
| | 3 | {5,6,27} | {21,22} | ({6},{6}) |
| | 4 | {7,8,28} | {23,24} | ({6},{6}) |
| $\mathcal{D}\left(r_1^{(3)}\right)$ | 1 | {1,2,25} | {1,2,25} | ({3},{3},{10}) |
| | 2 | {3,4,26} | {3,4,26} | ({3},{3},{10}) |
| | 3 | {5,6,27} | {5,6,27} | ({3},{3},{10}) |
| | 4 | {7,8,28} | {7,8,28} | ({3},{3},{10}) |
| $\mathcal{D}\left(r_1^{(5)}\right)$ | 1 | {9} | {1,2,3,4,5,6,7,8,25,26,27,28} | ({1,2},{1,2},{1,2},{1,2},{1,2},{1,2},{1,2},{1,2},{8,9},{8,9},{8,9},{8,9}) |
| | 2 | {10} | {1,2,3,4,5,6,7,8,25,26,27,28} | ({1,2},{1,2},{1,2},{1,2},{1,2},{1,2},{1,2},{1,2},{8,9},{8,9},{8,9},{8,9}) |
| | 3 | {11} | {1,2,3,4,5,6,7,8,25,26,27,28} | ({1,2},{1,2},{1,2},{1,2},{1,2},{1,2},{1,2},{1,2},{8,9},{8,9},{8,9},{8,9}) |
| | 4 | {12} | {1,2,3,4,5,6,7,8,25,26,27,28} | ({1,2},{1,2},{1,2},{1,2},{1,2},{1,2},{1,2},{1,2},{8,9},{8,9},{8,9},{8,9}) |
| | 5 | {13} | {1,2,3,4,5,6,7,8,25,26,27,28} | ({1,2},{1,2},{1,2},{1,2},{1,2},{1,2},{1,2},{1,2},{8,9},{8,9},{8,9},{8,9}) |
| | 6 | {14} | {1,2,3,4,5,6,7,8,25,26,27,28} | ({1,2},{1,2},{1,2},{1,2},{1,2},{1,2},{1,2},{1,2},{8,9},{8,9},{8,9},{8,9}) |
| | 7 | {15} | {1,2,3,4,5,6,7,8,25,26,27,28} | ({1,2},{1,2},{1,2},{1,2},{1,2},{1,2},{1,2},{1,2},{8,9},{8,9},{8,9},{8,9}) |
| | 8 | {16} | {1,2,3,4,5,6,7,8,25,26,27,28} | ({1,2},{1,2},{1,2},{1,2},{1,2},{1,2},{1,2},{1,2},{8,9},{8,9},{8,9},{8,9}) |
| $\mathcal{D}\left(r_2^{(5)}\right)$ | 1 | {9} | {1,2,3,4,5,6,7,8,25,26,27,28} | ({3},{3},{1,2},{1,2},{1,2},{1,2},{1,2},{1,2},{10},{8,9},{8,9},{8,9}) |
| | 2 | {10} | {1,2,3,4,5,6,7,8,25,26,27,28} | ({3},{3},{1,2},{1,2},{1,2},{1,2},{1,2},{1,2},{10},{8,9},{8,9},{8,9}) |
| | 3 | {11} | {1,2,3,4,5,6,7,8,25,26,27,28} | ({1,2},{1,2},{3},{3},{1,2},{1,2},{1,2},{1,2},{8,9},{10},{8,9},{8,9}) |
| | 4 | {12} | {1,2,3,4,5,6,7,8,25,26,27,28} | ({1,2},{1,2},{3},{3},{1,2},{1,2},{1,2},{1,2},{8,9},{10},{8,9},{8,9}) |
| | 5 | {13} | {1,2,3,4,5,6,7,8,25,26,27,28} | ({1,2},{1,2},{1,2},{1,2},{3},{3},{1,2},{1,2},{8,9},{8,9},{10},{8,9}) |
| | 6 | {14} | {1,2,3,4,5,6,7,8,25,26,27,28} | ({1,2},{1,2},{1,2},{1,2},{3},{3},{1,2},{1,2},{8,9},{8,9},{10},{8,9}) |
| | 7 | {15} | {1,2,3,4,5,6,7,8,25,26,27,28} | ({1,2},{1,2},{1,2},{1,2},{1,2},{1,2},{3},{3},{8,9},{8,9},{8,9},{10}) |
| | 8 | {16} | {1,2,3,4,5,6,7,8,25,26,27,28} | ({1,2},{1,2},{1,2},{1,2},{1,2},{1,2},{3},{3},{8,9},{8,9},{8,9},{10}) |
| $\mathcal{D}\left(r_3^{(5)}\right)$ | 1 | {9} | {1,2,3,4,5,6,7,8,25,26,27,28} | ({1,2},{1,2},{3},{3},{1,2},{1,2},{1,2},{1,2},{8,9},{10},{8,9},{8,9}) |
| | 2 | {9} | {1,2,3,4,5,6,7,8,25,26,27,28} | ({1,2},{1,2},{1,2},{1,2},{1,2},{1,2},{3},{3},{8,9},{8,9},{8,9},{10}) |
| | 3 | {10} | {1,2,3,4,5,6,7,8,25,26,27,28} | ({1,2},{1,2},{3},{3},{1,2},{1,2},{1,2},{1,2},{8,9},{10},{8,9},{8,9}) |
| | 4 | {10} | {1,2,3,4,5,6,7,8,25,26,27,28} | ({1,2},{1,2},{1,2},{1,2},{1,2},{1,2},{3},{3},{8,9},{8,9},{8,9},{10}) |
| | 5 | {11} | {1,2,3,4,5,6,7,8,25,26,27,28} | ({1,2},{1,2},{1,2},{1,2},{3},{3},{1,2},{1,2},{8,9},{8,9},{10},{8,9}) |
| | 6 | {11} | {1,2,3,4,5,6,7,8,25,26,27,28} | ({3},{3},{1,2},{1,2},{1,2},{1,2},{1,2},{1,2},{10},{8,9},{8,9},{8,9}) |
| | 7 | {12} | {1,2,3,4,5,6,7,8,25,26,27,28} | ({1,2},{1,2},{1,2},{1,2},{3},{3},{1,2},{1,2},{8,9},{8,9},{10},{8,9}) |
| | 8 | {12} | {1,2,3,4,5,6,7,8,25,26,27,28} | ({3},{3},{1,2},{1,2},{1,2},{1,2},{1,2},{1,2},{10},{8,9},{8,9},{8,9}) |
| | 9 | {13} | {1,2,3,4,5,6,7,8,25,26,27,28} | ({1,2},{1,2},{1,2},{1,2},{1,2},{1,2},{3},{3},{8,9},{8,9},{8,9},{10}) |
| | 10 | {13} | {1,2,3,4,5,6,7,8,25,26,27,28} | ({1,2},{1,2},{3},{3},{1,2},{1,2},{1,2},{1,2},{8,9},{10},{8,9},{8,9}) |
| | 11 | {14} | {1,2,3,4,5,6,7,8,25,26,27,28} | ({1,2},{1,2},{1,2},{1,2},{1,2},{1,2},{3},{3},{8,9},{8,9},{8,9},{10}) |
| | 12 | {14} | {1,2,3,4,5,6,7,8,25,26,27,28} | ({1,2},{1,2},{3},{3},{1,2},{1,2},{1,2},{1,2},{8,9},{10},{8,9},{8,9}) |
| | 13 | {15} | {1,2,3,4,5,6,7,8,25,26,27,28} | ({3},{3},{1,2},{1,2},{1,2},{1,2},{1,2},{1,2},{10},{8,9},{8,9},{8,9}) |
| | 14 | {15} | {1,2,3,4,5,6,7,8,25,26,27,28} | ({1,2},{1,2},{1,2},{1,2},{3},{3},{1,2},{1,2},{8,9},{8,9},{10},{8,9}) |
| | 15 | {16} | {1,2,3,4,5,6,7,8,25,26,27,28} | ({3},{3},{1,2},{1,2},{1,2},{1,2},{1,2},{1,2},{10},{8,9},{8,9},{8,9}) |
| | 16 | {16} | {1,2,3,4,5,6,7,8,25,26,27,28} | ({1,2},{1,2},{1,2},{1,2},{3},{3},{1,2},{1,2},{8,9},{8,9},{10},{8,9}) |
| $\mathcal{D}\left(r_4^{(5)}\right)$ | 1 | {9} | {1,2,3,4,5,6,7,8,25,26,27,28} | ({3},{3},{3},{3},{1,2},{1,2},{1,2},{1,2},{10},{10},{8,9},{8,9}) |
| | 2 | {9} | {1,2,3,4,5,6,7,8,25,26,27,28} | ({3},{3},{1,2},{1,2},{1,2},{1,2},{3},{3},{10},{8,9},{8,9},{10}) |
| | 3 | {10} | {1,2,3,4,5,6,7,8,25,26,27,28} | ({3},{3},{3},{3},{1,2},{1,2},{1,2},{1,2},{10},{10},{8,9},{8,9}) |
| | 4 | {10} | {1,2,3,4,5,6,7,8,25,26,27,28} | ({3},{3},{1,2},{1,2},{1,2},{1,2},{3},{3},{10},{8,9},{8,9},{10}) |
| | 5 | {11} | {1,2,3,4,5,6,7,8,25,26,27,28} | ({1,2},{1,2},{3},{3},{3},{3},{1,2},{1,2},{8,9},{10},{10},{8,9}) |
| | 6 | {11} | {1,2,3,4,5,6,7,8,25,26,27,28} | ({3},{3},{3},{3},{1,2},{1,2},{1,2},{1,2},{10},{10},{8,9},{8,9}) |
| | 7 | {12} | {1,2,3,4,5,6,7,8,25,26,27,28} | ({1,2},{1,2},{3},{3},{3},{3},{1,2},{1,2},{8,9},{10},{10},{8,9}) |
| | 8 | {12} | {1,2,3,4,5,6,7,8,25,26,27,28} | ({3},{3},{3},{3},{1,2},{1,2},{1,2},{1,2},{10},{10},{8,9},{8,9}) |
| | 9 | {13} | {1,2,3,4,5,6,7,8,25,26,27,28} | ({1,2},{1,2},{1,2},{1,2},{3},{3},{3},{3},{8,9},{8,9},{10},{10}) |
| | 10 | {13} | {1,2,3,4,5,6,7,8,25,26,27,28} | ({1,2},{1,2},{3},{3},{3},{3},{1,2},{1,2},{8,9},{10},{10},{8,9}) |
| | 11 | {14} | {1,2,3,4,5,6,7,8,25,26,27,28} | ({1,2},{1,2},{1,2},{1,2},{3},{3},{3},{3},{8,9},{8,9},{10},{10}) |
| | 12 | {14} | {1,2,3,4,5,6,7,8,25,26,27,28} | ({1,2},{1,2},{3},{3},{3},{3},{1,2},{1,2},{8,9},{10},{10},{8,9}) |
| | 13 | {15} | {1,2,3,4,5,6,7,8,25,26,27,28} | ({3},{3},{1,2},{1,2},{1,2},{1,2},{3},{3},{10},{8,9},{8,9},{10}) |
| | 14 | {15} | {1,2,3,4,5,6,7,8,25,26,27,28} | ({1,2},{1,2},{1,2},{1,2},{3},{3},{3},{3},{8,9},{8,9},{10},{10}) |
| | 15 | {16} | {1,2,3,4,5,6,7,8,25,26,27,28} | ({3},{3},{1,2},{1,2},{1,2},{1,2},{3},{3},{10},{8,9},{8,9},{10}) |
| | 16 | {16} | {1,2,3,4,5,6,7,8,25,26,27,28} | ({1,2},{1,2},{1,2},{1,2},{3},{3},{3},{3},{8,9},{8,9},{10},{10}) |

…

…

| | | | | |
|---|---|---|---|---|
| | 1 | {9} | {1,2,3,4,5,6,7,8,25,26,27,28} | ({1,2},{1,2},{3},{3},{1,2},{1,2},{3},{3},{8,9},{10},{8,9},{10}) |
| | 2 | {10} | {1,2,3,4,5,6,7,8,25,26,27,28} | ({1,2},{1,2},{3},{3},{1,2},{1,2},{3},{3},{8,9},{10},{8,9},{10}) |
| | 3 | {11} | {1,2,3,4,5,6,7,8,25,26,27,28} | ({3},{3},{1,2},{1,2},{3},{3},{1,2},{1,2},{10},{8,9},{10},{8,9}) |
| $\mathcal{D}\left(r_5^{(5)}\right)$ | 4 | {12} | {1,2,3,4,5,6,7,8,25,26,27,28} | ({3},{3},{1,2},{1,2},{3},{3},{1,2},{1,2},{10},{8,9},{10},{8,9}) |
| | 5 | {13} | {1,2,3,4,5,6,7,8,25,26,27,28} | ({1,2},{1,2},{3},{3},{1,2},{1,2},{3},{3},{8,9},{10},{8,9},{10}) |
| | 6 | {14} | {1,2,3,4,5,6,7,8,25,26,27,28} | ({1,2},{1,2},{3},{3},{1,2},{1,2},{3},{3},{8,9},{10},{8,9},{10}) |
| | 7 | {15} | {1,2,3,4,5,6,7,8,25,26,27,28} | ({3},{3},{1,2},{1,2},{3},{3},{1,2},{1,2},{10},{8,9},{10},{8,9}) |
| | 8 | {16} | {1,2,3,4,5,6,7,8,25,26,27,28} | ({3},{3},{1,2},{1,2},{3},{3},{1,2},{1,2},{10},{8,9},{10},{8,9}) |
| | 1 | {9} | {1,2,3,4,5,6,7,8,25,26,27,28} | ({3},{3},{3},{3},{1,2},{1,2},{3},{3},{10},{10},{8,9},{10}) |
| | 2 | {10} | {1,2,3,4,5,6,7,8,25,26,27,28} | ({3},{3},{3},{3},{1,2},{1,2},{3},{3},{10},{10},{8,9},{10}) |
| | 3 | {11} | {1,2,3,4,5,6,7,8,25,26,27,28} | ({3},{3},{3},{3},{3},{3},{1,2},{1,2},{10},{10},{10},{8,9}) |
| $\mathcal{D}\left(r_6^{(5)}\right)$ | 4 | {12} | {1,2,3,4,5,6,7,8,25,26,27,28} | ({3},{3},{3},{3},{3},{3},{1,2},{1,2},{10},{10},{10},{8,9}) |
| | 5 | {13} | {1,2,3,4,5,6,7,8,25,26,27,28} | ({1,2},{1,2},{3},{3},{3},{3},{3},{3},{8,9},{10},{10},{10}) |
| | 6 | {14} | {1,2,3,4,5,6,7,8,25,26,27,28} | ({1,2},{1,2},{3},{3},{3},{3},{3},{3},{8,9},{10},{10},{10}) |
| | 7 | {15} | {1,2,3,4,5,6,7,8,25,26,27,28} | ({3},{3},{1,2},{1,2},{3},{3},{3},{3},{10},{8,9},{10},{10}) |
| | 8 | {16} | {1,2,3,4,5,6,7,8,25,26,27,28} | ({3},{3},{1,2},{1,2},{3},{3},{3},{3},{10},{8,9},{10},{10}) |
| | 1 | {9} | {1,2,3,4,5,6,7,8,25,26,27,28} | ({1,2},{1,2},{1,2},{1,2},{3},{3},{1,2},{1,2},{8,9},{8,9},{10},{8,9}) |
| | 2 | {10} | {1,2,3,4,5,6,7,8,25,26,27,28} | ({1,2},{1,2},{1,2},{1,2},{3},{3},{1,2},{1,2},{8,9},{8,9},{10},{8,9}) |
| | 3 | {11} | {1,2,3,4,5,6,7,8,25,26,27,28} | ({1,2},{1,2},{1,2},{1,2},{1,2},{1,2},{3},{3},{8,9},{8,9},{8,9},{10}) |
| $\mathcal{D}\left(r_7^{(5)}\right)$ | 4 | {12} | {1,2,3,4,5,6,7,8,25,26,27,28} | ({1,2},{1,2},{1,2},{1,2},{1,2},{1,2},{3},{3},{8,9},{8,9},{8,9},{10}) |
| | 5 | {13} | {1,2,3,4,5,6,7,8,25,26,27,28} | ({3},{3},{1,2},{1,2},{1,2},{1,2},{1,2},{1,2},{10},{8,9},{8,9},{8,9}) |
| | 6 | {14} | {1,2,3,4,5,6,7,8,25,26,27,28} | ({3},{3},{1,2},{1,2},{1,2},{1,2},{1,2},{1,2},{10},{8,9},{8,9},{8,9}) |
| | 7 | {15} | {1,2,3,4,5,6,7,8,25,26,27,28} | ({1,2},{1,2},{3},{3},{1,2},{1,2},{1,2},{1,2},{8,9},{10},{8,9},{8,9}) |
| | 8 | {16} | {1,2,3,4,5,6,7,8,25,26,27,28} | ({1,2},{1,2},{3},{3},{1,2},{1,2},{1,2},{1,2},{8,9},{10},{8,9},{8,9}) |
| | 1 | {9} | {1,2,3,4,5,6,7,8,25,26,27,28} | ({3},{3},{1,2},{1,2},{3},{3},{1,2},{1,2},{10},{8,9},{10},{8,9}) |
| | 2 | {10} | {1,2,3,4,5,6,7,8,25,26,27,28} | ({3},{3},{1,2},{1,2},{3},{3},{1,2},{1,2},{10},{8,9},{10},{8,9}) |
| | 3 | {11} | {1,2,3,4,5,6,7,8,25,26,27,28} | ({1,2},{1,2},{3},{3},{1,2},{1,2},{3},{3},{8,9},{10},{8,9},{10}) |
| $\mathcal{D}\left(r_8^{(5)}\right)$ | 4 | {12} | {1,2,3,4,5,6,7,8,25,26,27,28} | ({1,2},{1,2},{3},{3},{1,2},{1,2},{3},{3},{8,9},{10},{8,9},{10}) |
| | 5 | {13} | {1,2,3,4,5,6,7,8,25,26,27,28} | ({3},{3},{1,2},{1,2},{3},{3},{1,2},{1,2},{10},{8,9},{10},{8,9}) |
| | 6 | {14} | {1,2,3,4,5,6,7,8,25,26,27,28} | ({3},{3},{1,2},{1,2},{3},{3},{1,2},{1,2},{10},{8,9},{10},{8,9}) |
| | 7 | {15} | {1,2,3,4,5,6,7,8,25,26,27,28} | ({1,2},{1,2},{3},{3},{1,2},{1,2},{3},{3},{8,9},{10},{8,9},{10}) |
| | 8 | {16} | {1,2,3,4,5,6,7,8,25,26,27,28} | ({1,2},{1,2},{3},{3},{1,2},{1,2},{3},{3},{8,9},{10},{8,9},{10}) |
| | 1 | {9} | {1,2,3,4,5,6,7,8,25,26,27,28} | ({1,2},{1,2},{3},{3},{3},{3},{1,2},{1,2},{8,9},{10},{10},{8,9}) |
| | 2 | {9} | {1,2,3,4,5,6,7,8,25,26,27,28} | ({1,2},{1,2},{1,2},{1,2},{3},{3},{3},{3},{8,9},{8,9},{10},{10}) |
| | 3 | {10} | {1,2,3,4,5,6,7,8,25,26,27,28} | ({1,2},{1,2},{3},{3},{3},{3},{1,2},{1,2},{8,9},{10},{10},{8,9}) |
| | 4 | {10} | {1,2,3,4,5,6,7,8,25,26,27,28} | ({1,2},{1,2},{1,2},{1,2},{3},{3},{3},{3},{8,9},{8,9},{10},{10}) |
| | 5 | {11} | {1,2,3,4,5,6,7,8,25,26,27,28} | ({1,2},{1,2},{1,2},{1,2},{3},{3},{3},{3},{8,9},{8,9},{10},{10}) |
| | 6 | {11} | {1,2,3,4,5,6,7,8,25,26,27,28} | ({3},{3},{1,2},{1,2},{1,2},{1,2},{3},{3},{10},{8,9},{8,9},{10}) |
| | 7 | {12} | {1,2,3,4,5,6,7,8,25,26,27,28} | ({1,2},{1,2},{1,2},{1,2},{3},{3},{3},{3},{8,9},{8,9},{10},{10}) |
| $\mathcal{D}\left(r_9^{(5)}\right)$ | 8 | {12} | {1,2,3,4,5,6,7,8,25,26,27,28} | ({3},{3},{1,2},{1,2},{1,2},{1,2},{3},{3},{10},{8,9},{8,9},{10}) |
| | 9 | {13} | {1,2,3,4,5,6,7,8,25,26,27,28} | ({3},{3},{1,2},{1,2},{1,2},{1,2},{3},{3},{10},{8,9},{8,9},{10}) |
| | 10 | {13} | {1,2,3,4,5,6,7,8,25,26,27,28} | ({3},{3},{3},{3},{1,2},{1,2},{1,2},{1,2},{10},{10},{8,9},{8,9}) |
| | 11 | {14} | {1,2,3,4,5,6,7,8,25,26,27,28} | ({3},{3},{1,2},{1,2},{1,2},{1,2},{3},{3},{10},{8,9},{8,9},{10}) |
| | 12 | {14} | {1,2,3,4,5,6,7,8,25,26,27,28} | ({3},{3},{3},{3},{1,2},{1,2},{1,2},{1,2},{10},{10},{8,9},{8,9}) |
| | 13 | {15} | {1,2,3,4,5,6,7,8,25,26,27,28} | ({3},{3},{3},{3},{1,2},{1,2},{1,2},{1,2},{10},{10},{8,9},{8,9}) |
| | 14 | {15} | {1,2,3,4,5,6,7,8,25,26,27,28} | ({1,2},{1,2},{3},{3},{3},{3},{1,2},{1,2},{8,9},{10},{10},{8,9}) |
| | 15 | {16} | {1,2,3,4,5,6,7,8,25,26,27,28} | ({3},{3},{3},{3},{1,2},{1,2},{1,2},{1,2},{10},{10},{8,9},{8,9}) |
| | 16 | {16} | {1,2,3,4,5,6,7,8,25,26,27,28} | ({1,2},{1,2},{3},{3},{3},{3},{1,2},{1,2},{8,9},{10},{10},{8,9}) |

…

…

| | | | | |
|---|---|---|---|---|
| $\mathcal{D}\left(r_{10}^{(5)}\right)$ | 1 | {9} | {1,2,3,4,5,6,7,8,25,26,27,28} | ({3},{3},{3},{3},{3},{3},{1,2},{1,2},{10},{10},{10},{8,9}) |
| | 2 | {9} | {1,2,3,4,5,6,7,8,25,26,27,28} | ({3},{3},{1,2},{1,2},{3},{3},{3},{3},{10},{8,9},{10},{10}) |
| | 3 | {10} | {1,2,3,4,5,6,7,8,25,26,27,28} | ({3},{3},{3},{3},{3},{3},{1,2},{1,2},{10},{10},{10},{8,9}) |
| | 4 | {10} | {1,2,3,4,5,6,7,8,25,26,27,28} | ({3},{3},{1,2},{1,2},{3},{3},{3},{3},{10},{8,9},{10},{10}) |
| | 5 | {11} | {1,2,3,4,5,6,7,8,25,26,27,28} | ({1,2},{1,2},{3},{3},{3},{3},{3},{3},{8,9},{10},{10},{10}) |
| | 6 | {11} | {1,2,3,4,5,6,7,8,25,26,27,28} | ({3},{3},{3},{3},{1,2},{1,2},{3},{3},{10},{10},{8,9},{10}) |
| | 7 | {12} | {1,2,3,4,5,6,7,8,25,26,27,28} | ({1,2},{1,2},{3},{3},{3},{3},{3},{3},{8,9},{10},{10},{10}) |
| | 8 | {12} | {1,2,3,4,5,6,7,8,25,26,27,28} | ({3},{3},{3},{3},{1,2},{1,2},{3},{3},{10},{10},{8,9},{10}) |
| | 9 | {13} | {1,2,3,4,5,6,7,8,25,26,27,28} | ({3},{3},{1,2},{1,2},{3},{3},{3},{3},{10},{8,9},{10},{10}) |
| | 10 | {13} | {1,2,3,4,5,6,7,8,25,26,27,28} | ({3},{3},{3},{3},{3},{3},{1,2},{1,2},{10},{10},{10},{8,9}) |
| | 11 | {14} | {1,2,3,4,5,6,7,8,25,26,27,28} | ({3},{3},{1,2},{1,2},{3},{3},{3},{3},{10},{8,9},{10},{10}) |
| | 12 | {14} | {1,2,3,4,5,6,7,8,25,26,27,28} | ({3},{3},{3},{3},{3},{3},{1,2},{1,2},{10},{10},{10},{8,9}) |
| | 13 | {15} | {1,2,3,4,5,6,7,8,25,26,27,28} | ({3},{3},{3},{3},{1,2},{1,2},{3},{3},{10},{10},{8,9},{10}) |
| | 14 | {15} | {1,2,3,4,5,6,7,8,25,26,27,28} | ({1,2},{1,2},{3},{3},{3},{3},{3},{3},{8,9},{10},{10},{10}) |
| | 15 | {16} | {1,2,3,4,5,6,7,8,25,26,27,28} | ({3},{3},{3},{3},{1,2},{1,2},{3},{3},{10},{10},{8,9},{10}) |
| | 16 | {16} | {1,2,3,4,5,6,7,8,25,26,27,28} | ({1,2},{1,2},{3},{3},{3},{3},{3},{3},{8,9},{10},{10},{10}) |
| $\mathcal{D}\left(r_{11}^{(5)}\right)$ | 1 | {9} | {1,2,3,4,5,6,7,8,25,26,27,28} | ({1,2},{1,2},{3},{3},{3},{3},{3},{3},{8,9},{10},{10},{10}) |
| | 2 | {10} | {1,2,3,4,5,6,7,8,25,26,27,28} | ({1,2},{1,2},{3},{3},{3},{3},{3},{3},{8,9},{10},{10},{10}) |
| | 3 | {11} | {1,2,3,4,5,6,7,8,25,26,27,28} | ({3},{3},{1,2},{1,2},{3},{3},{3},{3},{10},{8,9},{10},{10}) |
| | 4 | {12} | {1,2,3,4,5,6,7,8,25,26,27,28} | ({3},{3},{1,2},{1,2},{3},{3},{3},{3},{10},{8,9},{10},{10}) |
| | 5 | {13} | {1,2,3,4,5,6,7,8,25,26,27,28} | ({3},{3},{3},{3},{1,2},{1,2},{3},{3},{10},{10},{8,9},{10}) |
| | 6 | {14} | {1,2,3,4,5,6,7,8,25,26,27,28} | ({3},{3},{3},{3},{1,2},{1,2},{3},{3},{10},{10},{8,9},{10}) |
| | 7 | {15} | {1,2,3,4,5,6,7,8,25,26,27,28} | ({3},{3},{3},{3},{3},{3},{1,2},{1,2},{10},{10},{10},{8,9}) |
| | 8 | {16} | {1,2,3,4,5,6,7,8,25,26,27,28} | ({3},{3},{3},{3},{3},{3},{1,2},{1,2},{10},{10},{10},{8,9}) |
| $\mathcal{D}\left(r_{12}^{(5)}\right)$ | 1 | {9} | {1,2,3,4,5,6,7,8,25,26,27,28} | ({3},{3},{3},{3},{3},{3},{3},{3},{10},{10},{10},{10}) |
| | 2 | {10} | {1,2,3,4,5,6,7,8,25,26,27,28} | ({3},{3},{3},{3},{3},{3},{3},{3},{10},{10},{10},{10}) |
| | 3 | {11} | {1,2,3,4,5,6,7,8,25,26,27,28} | ({3},{3},{3},{3},{3},{3},{3},{3},{10},{10},{10},{10}) |
| | 4 | {12} | {1,2,3,4,5,6,7,8,25,26,27,28} | ({3},{3},{3},{3},{3},{3},{3},{3},{10},{10},{10},{10}) |
| | 5 | {13} | {1,2,3,4,5,6,7,8,25,26,27,28} | ({3},{3},{3},{3},{3},{3},{3},{3},{10},{10},{10},{10}) |
| | 6 | {14} | {1,2,3,4,5,6,7,8,25,26,27,28} | ({3},{3},{3},{3},{3},{3},{3},{3},{10},{10},{10},{10}) |
| | 7 | {15} | {1,2,3,4,5,6,7,8,25,26,27,28} | ({3},{3},{3},{3},{3},{3},{3},{3},{10},{10},{10},{10}) |
| | 8 | {16} | {1,2,3,4,5,6,7,8,25,26,27,28} | ({3},{3},{3},{3},{3},{3},{3},{3},{10},{10},{10},{10}) |
| $\mathcal{D}\left(r_{1}^{(6)}\right)$ | 1 | {9} | {17,18} | ({6},{6}) |
| | 2 | {10} | {17,18} | ({6},{6}) |
| | 3 | {11} | {19,20} | ({6},{6}) |
| | 4 | {12} | {19,20} | ({6},{6}) |
| | 5 | {13} | {21,22} | ({6},{6}) |
| | 6 | {14} | {21,22} | ({6},{6}) |
| | 7 | {15} | {23,24} | ({6},{6}) |
| | 8 | {16} | {23,24} | ({6},{6}) |
| $\mathcal{D}\left(r_{2}^{(6)}\right)$ | 1 | {9} | {17,18} | ({7},{7}) |
| | 2 | {10} | {17,18} | ({7},{7}) |
| | 3 | {11} | {19,20} | ({7},{7}) |
| | 4 | {12} | {19,20} | ({7},{7}) |
| | 5 | {13} | {21,22} | ({7},{7}) |
| | 6 | {14} | {21,22} | ({7},{7}) |
| | 7 | {15} | {23,24} | ({7},{7}) |
| | 8 | {16} | {23,24} | ({7},{7}) |
| $\mathcal{D}\left(r_{1}^{(7)}\right)$ | 1 | {17,18} | {1,2,25} | ({1,2},{1,2},{8,9}) |
| | 2 | {19,20} | {3,4,26} | ({1,2},{1,2},{8,9}) |
| | 3 | {21,22} | {5,6,27} | ({1,2},{1,2},{8,9}) |
| | 4 | {23,24} | {7,8,28} | ({1,2},{1,2},{8,9}) |

…

…

| | | | | |
|---|---|---|---|---|
| $\mathcal{D}\left(r_1^{(8)}\right)$ | 1 | {17,18} | {9,10} | ({4},{4}) |
| | 2 | {19,20} | {11,12} | ({4},{4}) |
| | 3 | {21,22} | {13,14} | ({4},{4}) |
| | 4 | {23,24} | {15,16} | ({4},{4}) |
| $\mathcal{D}\left(r_2^{(8)}\right)$ | 1 | {17,18} | {9,10} | ({4},{5}) |
| | 2 | {17,18} | {9,10} | ({5},{4}) |
| | 3 | {19,20} | {11,12} | ({4},{5}) |
| | 4 | {19,20} | {11,12} | ({5},{4}) |
| | 5 | {21,22} | {13,14} | ({4},{5}) |
| | 6 | {21,22} | {13,14} | ({5},{4}) |
| | 7 | {23,24} | {15,16} | ({4},{5}) |
| | 8 | {23,24} | {15,16} | ({5},{4}) |
| $\mathcal{D}\left(r_3^{(8)}\right)$ | 1 | {17,18} | {9,10} | ({5},{5}) |
| | 2 | {19,20} | {11,12} | ({5},{5}) |
| | 3 | {21,22} | {13,14} | ({5},{5}) |
| | 4 | {23,24} | {15,16} | ({5},{5}) |

**S4.** MATLAB script to compare transition matrices associated with the reduced state-space graphs generated by Vernan for the 28 and the four component $Na^+$ channel complex models of Examples 2 and 3. Since the two models specify the same system, the generated transition matrices are identical. This code snippet also illustrates the use of some principle functions in the Vernan toolbox.

```
%% Load models
load Example1
F_1channel = Vernan(n,Q,rules,D,rho,finit);

load Example3
F4 = F; clear F;
T4 = double(T); clear T;
W4 = W; clear W;
M4 = double(M); clear M;

load Example2
F28 = F; clear F;
T28 = double(T); clear T;
W28 = W; clear W;
M28 = double(M); clear M;

C = 10*rand(24,1); %randomly assign values to the 24 model rate constants
                   %in the interval (0,10)

%% Reconstruct 28 component system states from 4 component system states
F4to28 = zeros(size(F4,1),28);
for i = 1:size(F4,1)
    for j = 1:4
        F4to28(i,(1:2) + 2*(j-1)) = F_1channel(F4(i,j),1:2);
        F4to28(i,(9:10) + 2*(j-1)) = F_1channel(F4(i,j),3:4);
        F4to28(i,(17:18) + 2*(j-1)) = F_1channel(F4(i,j),5:6);
        F4to28(i,25 + j-1) = F_1channel(F4(i,j),7);
    end
end

% reconstructed 28 component system states need to be canonized in the same
% way as the 28 component model
[Type,Tsize,Trelab] = IDComponentTypes(n,Q);
[~,~,Gtilde] = GenerateCharacteristicGraph(finit,Type,rules,D);
[d,v,e,c] = SetupNautyGraph_sparse(Gtilde);
[S,vstab] = CallNauty_sparse(d,v,e,c);
S = S(:,1:28);
[base,~,~,U] = SchreierSims(S,vstab);
%warning! the loop below requires 304590 iterations and will take some time
%(~tens of minutes) to run
for i = 1:size(F4,1)
    F4to28(i,:) = FindCanonRep(F4to28(i,:),d,v,e,c,base,U);
end

%% Encode and sort system states
[numdigits,startofdigit,endofdigit] = GenerateCodingScheme(finit,Tsize);
[~,ixF28] = sort(Encode(F28,Tsize,Trelab,numdigits,startofdigit, ...
              endofdigit));
[~,ixF4to28] = sort(Encode(F4to28,Tsize,Trelab,numdigits, ...
                      startofdigit,endofdigit));
F28 = F28(ixF28);
F4to28 = F4to28(ixF4to28);

%need to update the indices in T for the two models to account for the new
%indexing of system states
ixF28(ixF28) = 1:length(ixF28);
T28 = ixF28(T28);
ixF4to28(ixF4to28) = 1:length(ixF4to28);
```

```
T4 = ixF4to28(T4);

%% Construct transition matrices for the 28 and 4 component versions of
%  the model and test their equivalence
A28 = sparse(T28(:,2),T28(:,1),M28.*C(W28));
A4 = sparse(T4(:,2),T4(:,1),M4.*Cprime(W4,1).*C(Cprime(W4,2)));
if isequal(A28,A4)
    disp('Matrices are equal');
else
    disp('Matrices are not equal');
end
```